\documentclass[12pt]{article}
\usepackage{amsmath,amssymb}
\usepackage[english]{babel}
\usepackage[cp866]{inputenc}
\numberwithin{equation}{section}
\newcommand{\bq}{\begin{eqnarray}}
\newcommand{\eq}{\end{eqnarray}}
\newcommand{\be}{\begin{equation}}
\newcommand{\ee}{\end{equation}}
\newcommand{\ra}{\rightarrow}
\newcommand{\la}{\Lambda_Q}
\newcommand{\mph}{\mu_{\Phi}}
\newcommand{\ov}{\overline}
\newcommand{\lym}{\Lambda_{YM}}

\newcommand{\bo}{{\rm b_o}}
\newcommand{\nd}{{\ov N}_c}
\newcommand{\sq}{\textsf{Q}}

\newcommand{\sqt}{\textsf{Q}_2}
\newcommand{\oq}{\ov{\textsf{Q}}}

\newcommand{\oqt}{\ov{\textsf{Q}}_2}
\newcommand{\dq}{\textsf{q}}
\newcommand{\odq}{\ov{\textsf{q}}}
\newcommand{\bd}{{\rm\ov b}_{\rm o}}

\newcommand{\qq}{{\ov Q}Q}
\newcommand{\qqo}{{\ov Q}_1Q_1}
\newcommand{\qqt}{{\ov Q}_2Q_2}
\newcommand{\qql}{\langle\qq\rangle_L}
\newcommand{\qqs}{\langle\qq\rangle_S}
\newcommand{\mgl}{\mu_{{\rm gl},\,L}}
\newcommand{\mgs}{\mu_{{\rm gl},\,S}}
\newcommand{\mql}{m^{\rm pole}_{Q,\,L}}
\newcommand{\mqs}{m^{\rm pole}_{Q,\,S}}
\newcommand{\mg}{\mu_{\rm gl}}
\newcommand{\mgo}{\mu_{\rm gl, 1}}
\newcommand{\mgt}{\mu_{\rm gl, 2}}
\newcommand{\mo}{\mu_{\Phi,\rm o}}

\newcommand{\qo}{\mu_{q,1}}
\newcommand{\qop}{\mu^{\rm pole}_{q,1}}
\newcommand{\qt}{\mu_{q,2}}
\newcommand{\qtp}{\mu^{\rm pole}_{q,2}}
\newcommand{\mug}{{\ov\mu}_{\rm gl}}
\newcommand{\muo}{{\ov\mu}_{{\rm gl},\,1}}
\newcommand{\mut}{{\ov\mu}_{{\rm gl},\,2}}

\begin{document}
\begin{center}{\bf \large Mass spectra in $\mathbf{{\cal N}=1}$ SQCD with additional fields. II} \end{center}
\vspace{1cm}
\begin{center}\bf Victor L. Chernyak \end{center}
\begin{center}(e-mail: v.l.chernyak@inp.nsk.su) \end{center}
\begin{center} Budker Institut of Nuclear Physics SB RAS and Novosibirsk State University,\\ 630090 Novosibirsk, Russia
\end{center}
\vspace{1cm}

\begin{center}{\bf Abstract} \end{center}
\vspace{1cm}

This article continues arXiV:\,1205.0410 [hep-th]. Considered is the ${\cal N}=1$ SQCD-like theory with $SU(N_c)$ colors and $3N_c/2< N_F<2N_c$ flavors of light quarks $\,Q_i,{\ov Q}_j$, and with the additional $N^2_F$ colorless flavored fields $\Phi_{ij}$ with the large mass parameter $\mph\gg\la$. The mass spectra of this $\Phi$ - theory (and its dual variant, the $d\Phi$ - theory) are calculated at different values of $\mph/\la\gg 1$ within the dynamical scenario which implies that quarks can be in two different phases only\,: either this is the HQ (heavy quark) phase where they are confined, or they are higgsed at appropriate values of the lagrangian parameters. It is shown that at the left end of the conformal window, i.e. at $0<(2N_F-3N_c)/N_F\ll 1$, the mass spectra of the direct and dual theories are parametrically different.

\newpage
{\Large\bf Contents}\\

{\bf 1 \,\,\,  Introduction}\\
1.1 \,\,\, Direct $\mathbf{\Phi}$ - theory \hspace*{12.2cm}\,\,3\\
1.2 \,\,\, Dual $\,\mathbf{d\Phi}$ - theory \hspace*{12.15cm}\,\,6
\vspace{1mm}

{\bf 2 \,\,\,  Direct theory. Unbroken flavor symmetry}\\
2.1 \,\,\,  $\mathbf L$ - vacua \hspace*{13.7cm}\,\,8\\
2.2 \,\,\,  $\mathbf S$ - vacua \hspace*{13.76cm}\,\,9
\vspace{1mm}

{\bf 3\,\,\,\,\,  Dual theory. Unbroken flavor symmetry}\hspace*{0.7cm}\\
3.1 \,\,\,  $\mathbf L$ - vacua, $\,\,\mathbf \bd/N_F\ll 1$ \hspace*{11.25cm}\,\,11\\
3.2 \,\,\,  $\mathbf S$ - vacua, $\,\,\mathbf \bd/N_F\ll 1$ \hspace*{11.27cm}\,\,13
\vspace{1mm}

{\bf 4 \,\,\, Direct theory. Broken flavor symmetry. The region $\mathbf{\la\ll\mph\ll\mo}$}\\
4.1 \,\,\, $\mathbf L$ - type vacua \hspace*{12.75cm}\,\,15\\
4.2 \,\,\, ${\rm\bf br2}$ vacua \hspace*{13.55cm}\,\,15\\
4.3 \,\,\, {\bf Special} vacua, $\mathbf{n_1=\nd,\,n_2=N_c}$ \hspace*{9.15cm}\,\,17
\vspace{1mm}

{\bf 5 \,\,\, Dual theory. Broken flavor symmetry. The region $\mathbf{\la\ll\mph\ll\mo}$}\\
5.1 \,\,\, $\mathbf L$ - type vacua,  $\,\,\mathbf \bd/N_F\ll 1$ \hspace*{10.3cm}\,\,\,17\\
5.2 \,\, ${\rm\bf br2}$ vacua,   $\,\,\mathbf \bd/N_F=O(1)$ \hspace*{10.52cm}\,\, 17\\
5.3 \,\, ${\rm\bf br2}$ vacua,  $\,\,\mathbf \bd/N_F\ll 1$ \hspace*{11.1cm}\,\, 19\\
5.4 \,\, {\bf Special} vacua, $\mathbf{n_1=\nd,\, n_2=N_c}$ \hspace*{9.25cm} 20
\vspace{1mm}

{\bf 6 \,\,\,  Direct theory. Broken flavor symmetry. The region $\mathbf{\mo\ll\mph\ll\la^2/m_Q}$}\\
6.1 \,\,\,   $\rm\bf br1$ vacua,  \hspace*{13.45cm}\,\,20\\
6.2 \,\,\,  $\rm \bf br2$ and {\bf special} vacua \hspace*{11.3cm}\,\,23
\vspace{1mm}

{\bf 7 \,\,\, Dual theory. Broken flavor symmetry. The region $\mathbf{\mo\ll\mph\ll\la^2/m_Q}$}\\
7.1 \,\,\,   $\rm\bf br1$ vacua,  $\,\,\mathbf \bd/N_F\ll 1$\hspace*{11.35cm}\,\,23\\
7.2 \,\,  $\rm \bf br2$ and {\bf special} vacua \hspace*{11.35cm}\,\,25
\vspace{1mm}

{\bf 8 \,\,\, Conclusions}\hspace*{12.6cm}\,\,  26\\
\vspace{1mm}

{\bf \,\,\, References}\hspace*{13.55cm}\,\,  26\\

\newpage

\section{Introduction}

We continue in this paper our study of mass spectra in the $\mathbf \Phi$ - theory (and in its dual variant $ d\mathbf \Phi$) that was started in \cite{ch1}. The difference with \cite{ch1} is that we calculate here mass spectra at
$3N_c/2<N_F<2N_c$ within the dynamical scenario $\#2$.
\footnote{\,
The mass spectra of the direct $\mathbf \Phi$ - theory at $0<N_F<N_c$ are the same in both scenarios $\#1$ and $\#2$,
see section 2 in \cite{ch1}.
}
We recall, see \cite{ch2}, that this scenario implies that quarks can be in the two different phases only. They are either in the heavy quark (HQ) phase where they are confined or they are higgsed (the Higgs phase), under appropriate conditions.

For a reader convenience, we recall below in this section the definitions of $\mathbf \Phi$ and $ d\mathbf \Phi$ - theories and some their most general properties (see sections 1 and 3 and the appendix A in \cite{ch1} for more details).\\

\hspace{1cm} {\bf 1.1.\,\, Direct $\mathbf \Phi$ - theory}\\

The field content of this direct ${\cal N}=1\,\,$ $\mathbf \Phi$ - theory includes $SU(N_c)$ gluons and $3N_c/2< N_F<2N_c$ flavors of light quarks ${\ov Q}_j, Q_i$. Besides, there is $N^2_F$ colorless but flavored fields $\Phi_{ji}$ (fions).

The Lagrangian at scales $\mu\leq\la$ is taken in the form
\bq
K={\rm Tr}\,\Bigl (\Phi^\dagger \Phi\Bigr )+{\rm Tr}\Bigl (\,Q^\dagger Q+(Q\ra {\ov Q})\,\Bigr )\,,
\quad W=-\frac{2\pi}{\alpha(\mu,\la)} S+W_{\Phi}+{\rm Tr}\,{\ov Q}(m_Q-\Phi) Q\,,\nonumber
\eq
\bq
W_{\Phi}=\frac{\mph}{2}\Biggl [{\rm Tr}\,(\Phi^2)-\frac{1}{\nd}\Bigl ({\rm Tr}\,\Phi\Bigr )^2\Biggr ]\,.
\eq
Here\,: $\mph$ and $m_Q$  are the mass parameters, $S=-W^{a}_{\beta}W^{a,\,\beta}/32\pi^2$ where $W^a_{\beta}$ is the gauge field strength, $a=1...N_c^2-1,\, \beta=1,2$,\, $\alpha(\mu,\la)=g^2(\mu,\la)/4\pi$ is the gauge coupling with its scale factor $\la$,\, $\nd=N_F-N_c$\,, the exponents with gluons in the quark Kahler term are implied here and below.
{\it This normalization of fields at $\mu=\la$ will be used everywhere below}. Besides, the perturbative NSVZ $\beta$-function for massless SUSY theories \cite{NSVZ1,NSVZ2} will be used everywhere in this paper.

Therefore, the $\mathbf \Phi$-theory we deal with in this paper has the parameters\,:\, $N_c\,,\,3N_c/2<N_F<2N_c\,
,\,\mph$,\, $\la,\, m_Q$, with the {\it strong hierarchies} $\mph\gg\la\gg m_Q$. The mass parameter $\mph$ will be varied while $m_Q$ and $\la$ will stay intact.\\

The Konishi anomalies \cite{Konishi} for the $i$-th flavor look as (${\it i}=1\, ...\, N_F$)
\bq
\langle\Phi_{ij}\rangle=\delta_{ij}\langle\Phi_{i}\rangle\,,\quad\langle\Phi_{i}\rangle\langle\frac{\partial W_{\Phi}}{\partial \Phi_{i}}\rangle=0\,,\quad
\langle m_{Q,i}^{\rm tot}\rangle\langle {\ov Q}_i Q_i\rangle=\langle S\rangle\,,\quad \langle m_{Q,\,i}^{\rm tot}\rangle=m_Q-\langle\Phi_{i}\rangle\,, \nonumber
\eq
\bq
\langle\Phi_{ij}\rangle=\frac{1}{\mph}\Biggl ( \langle{\ov Q}_j Q_i \rangle-\delta_{ji}\frac{1}{N_c}{\rm Tr}\,\langle\qq\rangle\Biggr )\,,\quad \langle{\ov Q}_j Q_i \rangle=\delta_{ji}\langle{\ov Q}_i Q_i \rangle\,,
\eq
and $\langle m_{Q,i}^{\rm tot}\rangle$ is the value of the quark running mass at $\mu=\la$.

At all scales $\mu<\la$ until the field $\Phi$ remains too heavy and non-dynamical, i.e. until its perturbative running mass $\mu_{\Phi}^{\rm pert}(\mu)>\mu$, it can be integrated out and the Lagrangian takes the form
\bq
K=z_Q(\la,\mu){\rm Tr}\Bigl (Q^\dagger Q+Q\ra {\ov Q}\Bigr ),\quad W=-\frac{2\pi}{\alpha(\mu,\la)}S+W_Q\,, \nonumber
\eq
\bq
W_Q=m_Q{\rm Tr}({\ov Q} Q)-\frac{1}{2\mph}\Biggl ({\rm Tr}\,({\ov Q}Q)^2-\frac{1}{N_c}\Bigl({\rm Tr}\,{\ov Q} Q \Bigr)^2 \Biggr ).
\eq

The Konishi anomalies for the Lagrangian (1.3) look as
\bq
\langle S\rangle=\langle {\ov Q}_i\frac{\partial W_Q}{\partial {\ov Q}_i}\rangle=m_Q\langle {\ov Q}_i Q_i \rangle-
\frac{1}{\mph}\Biggl (\sum_j\langle{\ov Q}_i Q_j\rangle\langle{\ov Q}_j Q_i\rangle-\frac{1}{N_c}\langle{\ov Q}_i Q_i\rangle\langle {\rm Tr}\,{\ov Q} Q \rangle\Biggr )=\nonumber
\eq
\bq
=\langle {\ov Q}_i Q_i \rangle\Biggl [\,m_Q-\frac{1}{\mph}\Biggl ( \langle {\ov Q}_i Q_i \rangle-\frac{1}{N_c}
\langle {\rm Tr}\,{\ov Q} Q \rangle\Biggr )  \Biggr ]\,,\quad i=1\,...\, N_F\,,\quad
\langle S\rangle=\langle\frac{\lambda\lambda}{32\pi^2}\rangle\,,
\eq
\bq
0=\langle {\ov Q}_i\frac{\partial W_Q}{\partial {\ov Q}_i}-{\ov Q}_j\frac{\partial W_Q}{\partial {\ov Q}_j}\rangle=\langle
{\ov Q}_i Q_i -{\ov Q}_j Q_j\rangle \Biggl [\,m_Q-\frac{1}{\mph}\Biggl ( \langle {\ov Q}_i Q_i+{\ov Q}_j Q_j\rangle
-\frac{1}{N_c}\langle {\rm Tr}\,{\ov Q} Q \rangle\Biggr )  \Biggr ]\,.\nonumber
\eq
It is most easily seen from (1.4) that there are only two types of vacua\,: a) the vacua with the unbroken flavor symmetry,\, b) the vacua with the spontaneously broken flavor symmetry, and the breaking is of the type $U(N_F)\ra U(n_1)
\times U(n_2)$ only. In these vacua
\bq
\langle {\ov Q}_1 Q_1+{\ov Q}_2 Q_2\rangle-\frac{1}{N_c}{\rm Tr}\,\langle {\ov Q} Q\rangle=m_Q\mph,\quad
\langle S\rangle=\frac{1}{\mph}\langle {\ov Q}_1 Q_1 \rangle\langle {\ov Q}_2 Q_2 \rangle,\quad
\langle {\ov Q}_1 Q_1 \rangle\neq\langle {\ov Q}_2 Q_2 \rangle\,,\nonumber
\eq
\bq
\langle m^{\rm tot}_{Q,1}\rangle=m_Q-\langle\Phi_1\rangle=\frac{\langle{\ov Q}_2 Q_2\rangle}{\mph},\quad \langle m^{\rm tot}_{Q,2}\rangle=m_Q-\langle\Phi_2\rangle=\frac{\langle{\ov Q}_1 Q_1\rangle}{\mph}\,.
\eq
\vspace*{2mm}

The values of quark condensates $\langle{\ov Q}_j Q_i\rangle=\delta_{ij}\langle{\ov Q}_i Q_i\rangle$ in different vacua
look as follows, see section 3 in \cite{ch1}.\\

\hspace{1cm} {\bf 1.1.1\,\, The region\,\,} $\la\ll\mph\ll\mo\,,\,\,\,\mo=\la\Bigl (\la/m_Q\Bigr )^{(2N_c-N_F)/N_c}$
\vspace*{2mm}

a) Unbroken flavor symmetry, $(2N_c-N_F)$\,\, L - vacua
\bq
\langle\qq\rangle_L\sim \la^2\Biggl (\frac{\la}{\mph}\Biggr )^{\frac{\nd}{2N_c-N_F}}\ll \la^2\,.
\eq

b) Unbroken flavor symmetry, $\nd=(N_F-N_c)$\,\, S - vacua
\bq
\langle\qq\rangle_S\simeq -\frac{\nd}{N_c}\, m_Q\mph\ll\la^2\,.
\eq

c) Broken flavor symmetry,\,  $U(N_F)\ra U(n_1)\times U(n_2),\, n_1\leq [N_F/2]$. In these, there are $n_1$ equal condensates $\langle\qqo\rangle$ and $n_2\geq n_1$ equal condensates $\langle\qqt\rangle\neq\langle\qqo\rangle$.
At $n_2\lessgtr N_c$, including $n_1=n_2=N_F/2$ for even $N_F$ but excluding $n_2=N_c$\,,
there are $(2N_c-N_F){\ov C}^{\,n_1}_{N_F}$ L - type vacua with
\bq
(1-\frac{n_1}{N_c})\langle\qqo\rangle_{Lt}\simeq -(1-\frac{n_2}{N_c})\langle\qqt\rangle_{Lt}\sim \la^2\Biggl (\frac{\la}{\mph}\Biggr )^{\frac{\nd}{2N_c-N_F}},
\eq
\bq
N_F=n_1+n_2\,,\quad n_1\leq [N_F/2]\,,\quad n_2\geq [N_F/2]\,,\quad C^{\,n_1}_{N_F}=\frac{N_F\,!}{n_1\,!\, n_2\,!}\,,
\nonumber
\eq
${\ov C}^{\,n_1}_{N_F}$ differ from the standard $C^{\,n_1}_{N_F}$ only by ${\ov C}^{\,n_1={\rm k}}_{N_F=2{\rm k}}=
C^{\,n_1={\rm k}}_{N_F=2{\rm k}}/2$.\\

d)  Broken flavor symmetry.\, At $n_2>N_c$ there are also $(n_2-N_c)C^{\,n_1}_{N_F}$ $\rm br2$ - vacua with
\bq
\langle\qqt\rangle_{\rm br2}\simeq \frac{N_c}{N_c-n_2}\, m_Q\mph\,,\quad \langle\qqo\rangle_{\rm br2}\sim \la^2\Bigl (\frac{\mph}{\la}\Bigr )^{\frac{n_2}{n_2-N_c}}\Bigl (\frac{m_Q}{\la}\Bigr )^{\frac{N_c-n_1}{n_2-N_c}}\,,
\eq
\bq
\frac{\langle\qqo\rangle_{\rm br2}}{\langle\qqt\rangle_{\rm br2}}\sim \Bigl (\frac{\mph}{\mo}\Bigr )^{\frac{N_c}{n_2-N_c}}\ll 1.\nonumber
\eq
\vspace*{2mm}

e)  Broken flavor symmetry.\, At $n_1=\nd,\, n_2=N_c$ there are $(2N_c-N_F) C^{\,n_1=\nd}_{N_F}$ "special" vacua with
\bq
\langle\qqo\rangle_{\rm spec}=\frac{N_c}{2N_c-N_F}\, m_Q\mph\,,\quad \langle\qqt\rangle_{\rm spec}\sim \la^2
\Bigl (\frac{\la}{\mph}\Bigr )^{\frac{\nd}{2N_c-N_F}}\,,
\eq
\bq
\frac{\langle\qqo\rangle_{\rm br2}}{\langle\qqt\rangle_{\rm br2}}\sim \Bigl (\frac{\mph}{\mo}\Bigr )^{\frac{N_c}{2N_c-N_F}}\ll 1\,.\nonumber
\eq

\hspace{1cm} {\bf 1.1.2\,\, The region\,\,} $\mo\ll\mph\ll\la^2/m_Q\,,\,\,\,\mo=\la\Bigl (\la/m_Q\Bigr )^{(2N_c-N_F)/N_c}$
\vspace*{2mm}

a) Unbroken flavor symmetry, $N_c$\,\, SQCD - vacua
\bq
\langle\qq\rangle_{SQCD}\sim\la^2\Biggl (\frac{m_Q}{\la}\Biggr )^{\nd/N_c}\,.
\eq

b) Broken flavor symmetry.\, At all values of $n_2\lessgtr N_c$, including $n_1=n_2=N_F/2$ at even $N_F$ and the "special" vacua with $n_1=\nd,\, n_2=N_c$, there are $(N_c-n_1){\ov C}^{\,n_1}_{N_F}$ $\rm br1$ - vacua with
\bq
\langle\qqo\rangle_{\rm br1}\simeq\frac{N_c}{N_c-n_1}\, m_Q\mph\,,\quad \langle\qqt\rangle_{\rm br1}\sim \la^2\Bigl (\frac{\la}{\mph}\Bigr )^{\frac{n_1}{N_c-n_1}}\Bigl (\frac{\la}{m_Q}\Bigr )^{\frac{N_c-n_2}{N_c-n_1}}\,,
\eq
\bq
\frac{\langle\qqt\rangle_{\rm br1}}{\langle\qqo\rangle_{\rm br1}}\sim \Bigl (\frac{\mo}{\mph}\Bigr )^{\frac{N_c}{N_c-n_1}}\ll 1\,.\nonumber
\eq

c) Broken flavor symmetry.\, At $n_2<N_c$, including $n_1=n_2=N_F/2$ at even $N_F$, there are $(N_c-n_2){\ov C}^{\,n_2}_{N_F}=(N_c-n_2){\ov C}^{\,n_1}_
{N_F}$\,\, $\rm br2$ - vacua with
\bq
\langle\qqt\rangle_{\rm br2}\simeq\frac{N_c}{N_c-n_2}\, m_Q\mph\,,\quad \langle\qqo\rangle_{\rm br2}\sim \la^2\Bigl (\frac{\la}{\mph}\Bigr )^{\frac{n_2}{N_c-n_2}}\Bigl (\frac{\la}{m_Q}\Bigr )^{\frac{N_c-n_1}{N_c-n_2}}\,,
\eq
\bq
\frac{\langle\qqo\rangle_{\rm br2}}{\langle\qqt\rangle_{\rm br2}}\sim \Bigl (\frac{\mo}{\mph}\Bigr )^{\frac{N_c}{N_c-n_2}}\ll 1\,.\nonumber
\eq

The perturbative running mass of fions $\mu^{\rm pert}_{\Phi}(\mu)=\mph/z_{\Phi}(\la,\mu\ll\la)\ll\mph$ decreases with
the diminishing scale $\mu\ll\la$ because $z_{\Phi}(\la,\mu\ll\la)\gg 1$. Hence, if nothing prevents, the fions become effectively massless and dynamically relevant at scales $\mu<\mu^{\rm conf}_o$, see section 4 in \cite{ch1},
\bq
\mu^{\rm conf}_o\sim\la\Bigl (\frac{\la}{\mph}\Bigr )^{N_F/3(2N_c-N_F)}\,.
\eq

The RG evolution of $\mu^{\rm pert}_{\Phi}(\mu)=\mph/z_{\Phi}(\la,\mu)$ becomes frozen at $\mu<\mu_H$, where $\mu_H<\la$ is the largest mass in the quark-gluon sector, $\mu^{\rm pert}_{\Phi}(\mu<\mu_H)=\mph/z_{\Phi}(\la,\mu=\mu_H)$
(within the dynamical scenario $\#2$ used in this paper, this $\mu_H$ may be either the quark pole mass $m_Q^{\rm pole}$ or the gluon mass $\mu_{\rm gl}$ if quarks are higgsed). Hence, if $\mu_H\gg\mu^{\rm conf}_o$, then fions remain too heavy and dynamically irrelevant at all scales $\mu<\la$. But if $\mu_H\ll\mu^{\rm conf}_o$, then there is a pole in the fion propagator at the momentum $p=\mu^{\rm conf}_o$, so that there appears the second generation of fions with the pole masses $\mu_2^{\rm pole}(\Phi)=\mu^{\rm conf}_o$. Moreover, because the limiting low energy value of the perturbative running mass of fions is $\mu^{\rm pert}(\Phi)=\mph/z_{\Phi}(\la,\mu_H)\ll\mu_H$ at $\mu_H\ll\mu^{\rm conf}_o$, and if this $\mu^{\rm pert}(\Phi)$ is the largest contribution to the fion mass, then there is also a pole in the fion propagator at the momentum $p=\mu^{\rm pert}(\Phi)$, so that there appears the third generation of fions with the pole masses $\mu_3^{\rm pole}(\Phi)=\mph/z_{\Phi}(\la,\mu_H)\ll\mu_H$.

In vacua with broken flavor symmetry the above reasonings have to be applied separately to $\Phi_{11},\, \Phi_{22}$ and
$\Phi_{12}+\Phi_{21}$.\\

\vspace*{2mm}

\hspace*{1cm} {\bf 1.2.\,\, Dual $\mathbf {d\Phi}$ - theory}\\

In parallel with the direct $\mathbf\Phi$ - theory we consider at $3N_c/2<N_F<2N_c$ also the Seiberg dual variant
\cite{S1, S2} (the $d\mathbf\Phi$ - theory), with the dual Lagrangian at $\mu=\Lambda_q$ \cite{ch1}
\bq
K={\rm Tr}\,\Bigl (\Phi^\dagger \Phi\Bigr )+ {\rm Tr}\Bigl ( q^\dagger q + (q\ra\ov q)\, \Bigr )+{\rm Tr}\, \Bigl (\frac{M^{\dagger}M}{\mu_2^2}\Bigr )\,,\quad
W=\, -\,\frac{2\pi}{\ov \alpha(\mu=\Lambda_q)}\, {\ov s}+{\ov W}_M+W_q\,,\nonumber
\eq
\bq
{\ov W}_M=\frac{\mph}{2}\Biggl [{\rm Tr}\,(\Phi^2)-\frac{1}{\nd}\Bigl ({\rm Tr}\,\Phi\Bigr )^2\Biggr ]+ {\rm Tr}\,
M(m_Q-\Phi),\quad W_q= -\,\frac{1}{\mu_1}\,\rm {Tr} \Bigl ({\ov q}\,M\, q \Bigr )\,.
\eq
Here\,:\, the number of dual colors is ${\ov N}_c=(N_F-N_c),\, \bd=3\nd-N_F$, and $M_{ij}$ are the $N_F^2$ elementary mion fields, ${\ov a}(\mu)=\nd{\ov \alpha}(\mu)/2\pi=\nd{\ov g}^2(\mu)/8\pi^2$ is the dual running gauge coupling,\,\,${\ov s}=-{\rm \ov w}^{b}_{\beta}{\rm \ov w}^{b,\,\beta}/32\pi^2$,\,\,
${\rm \ov w}^b_{\beta}$ is the dual gluon field strength. The gluino condensates of the direct and dual theories are matched, $\langle{-\,\ov s}\rangle=\langle S\rangle=\lym^3$.

By definition, $\mu=\Lambda_q$ is such a scale that, at $0<\bd/\nd\ll 1$, the dual theory already entered sufficiently deep into the conformal regime, i.e. both the gauge and Yukawa couplings, ${\ov a}(\mu=\Lambda
_q)$ and $a_f(\mu=\Lambda_q)=\nd f^2(\mu=\Lambda_q)/8\pi^2$, are already close to their small fixed point values, $\ov\delta=({\ov a}_*-{\ov a}(\mu=\Lambda_q)/{\ov a}_*\ll 1$, and similarly for $a_f(\mu=\Lambda
_q)$.  We take $|\Lambda_q|=\la$ for simplicity (because this does not matter finally, see the appendix in \cite{ch2} for more details). The condensates $\langle M_{ji}(\mu=\la)\rangle=\langle{\ov Q}_j Q_i(\mu=\la)
\rangle$ can always be matched at $\mu=|\Lambda_q|=\la$, at the appropriate choice of $\mu_1$ in (1.15).

At $3/2<N_F/N_c<2$ this dual theory can be taken as UV free at $\mu\gg\la$ and this requires that its Yukawa coupling at $\mu=\la,\, f(\mu=\la)=\mu_2/\mu_1$, cannot be larger than its gauge coupling ${\ov g}(\mu=\la)$, i.e. $\mu_2/\mu_1\lesssim 1$. The same requirement to the value of the Yukawa coupling follows from the conformal behavior of this theory at $3/2<N_F/N_c<2$ and $\mu<\la$, i.e. $a_f(\mu=\la)\simeq a^*_f\sim {\ov a}_*=O(\bd/N_F)$. We consider below this dual theory at $\mu\leq \la$ only where it claims to be equivalent to the direct $\Phi$ - theory.

As was explained in \cite{ch2} (see section 4 and appendix therein, the fixed point value of the dual gauge coupling at $\bd/N_F\ll 1$ is ${\ov a}_*\simeq 7\bd/3\nd$\, \cite{KSV}), one has to take
\bq
\mu_1\equiv Z_q\la\,,\quad Z_q\sim\exp \Bigl\{-\frac{1}{3{\ov a}_*}\Bigr \}\sim\exp \Bigl\{-\frac{\nd}{7\bd}\Bigr \}\ll 1\,,\quad \frac{\bd}{N_F}\ll 1
\eq
to match the gluino condensates in the direct and dual theories at $0<\bd/N_F\ll 1$. The value of $Z_q$ in (1.16) is valid with the exponential accuracy only, i.e. powers of $\bd/\nd\ll 1$ are ignored and only powers of $Z_q$ are tracked explicitly. This exponential accuracy is sufficient for our purposes. Besides, the small number $\bd/\nd\ll 1$ {\it does not compete in any way} with the much smaller parameter $m_Q/\la\ll 1$, see \cite{ch2}. At $\bd/N_F=O(1)$ one has to replace $Z_q\ra 1$ in (1.16). Then, from $\mu_2\lesssim\mu_1$ and (1.16), we take $\mu_2=\mu_1=Z_q\la$ in (1.15) everywhere below.

The fields $\Phi$ remain always too heavy and dynamically irrelevant in this $d\mathbf\Phi$ - theory, so that they can be integrated out once and forever and, finally, we write the Lagrangian of the dual $d\mathbf\Phi$ theory at $\mu=\la$ in the form
\bq
K= {\rm Tr}\Bigl ( q^\dagger q +(q\ra\ov q) \Bigr )+{\rm Tr}\,\frac{M^{\dagger}M}{Z_q^2\la^2}\,,\quad
W=\, -\,\frac{2\pi}{\ov \alpha(\mu=\la)}\, {\ov s}+W_M+W_q\,,\nonumber
\eq
\bq
W_M=m_Q{\rm Tr}\,M -\frac{1}{2\mph}\Biggl [{\rm Tr}\, (M^2)- \frac{1}{N_c}({\rm Tr}\, M)^2 \Biggr ]\,,\quad
W_q= -\,\frac{1}{Z_q\la}\,\rm {Tr} \Bigl ({\ov q}\,M\, q \Bigr )\,.
\eq

From (1.17), the Konishi anomalies for the $i$-th flavor, ${\it i}=1\, ...\, N_F$, look here as
\bq
\langle M_i\rangle\langle N_i\rangle=Z_q\la\langle S\rangle\,,
\quad \frac{\langle N_i\rangle}{Z_q\la}=m_Q-\frac{1}{\mph}\Bigl (\langle M_i-\frac{1}{N_c}{\rm Tr}\,M \rangle\Bigr )=
\langle m_{Q,i}^{\rm tot}\rangle\,,
\eq
\bq
\langle N_i\rangle\equiv\langle{\ov q}_i q_i(\mu=\la)\rangle\,.\nonumber
\eq

In vacua with the broken flavor symmetry these can be rewritten as
\bq
\langle M_1+M_2\rangle-\frac{1}{N_c}{\rm Tr}\,\langle M\rangle=m_Q\mph,\quad
\langle S\rangle=\frac{1}{\mph}\langle M_1\rangle\langle M_2\rangle,\quad
\langle M_1\rangle\neq\langle M_2\rangle\,,\nonumber
\eq
\bq
\frac{\langle N_1\rangle}{Z_q\la}= \frac{\langle S\rangle}{\langle M_{1}\rangle}=\frac{\langle M_{2}\rangle}{\mph}=m_Q-\frac{1}{\mph}\Bigl (\langle M_{1}\rangle-\frac{1}{N_c}{\rm Tr}\,\langle M\rangle \Bigr )=\langle m^{\rm tot}_{Q,1}\rangle\,,
\eq
\bq
\frac{\langle N_2\rangle}{Z_q\la}=\frac{\langle S\rangle}{\langle M_{2}\rangle}=\frac{\langle M_{1}\rangle}{\mph}=m_Q-\frac{1}{\mph}\Bigl (\langle M_{2}\rangle-\frac{1}{N_c}{\rm Tr}\,\langle M\rangle \Bigr )=\langle m^{\rm tot}_{Q,2}\rangle\,.\nonumber
\eq

The multiplicities of vacua are the same in the direct and dual theories \cite{ch1}.\\

Our purpose in this paper is to calculate mass spectra in the two above theories, $\mathbf\Phi$ and $\mathbf{d\Phi}$. At present, unfortunately, to calculate the mass spectra in ${\cal N}=1$ SQCD-like theories one has to rely on a definite dynamical scenario. Two different scenarios, $\#1$ and $\#2$, have been considered in \cite{ch3, ch4} and \cite{ch2} and the mass spectra were calculated in the standard direct ${\cal N}=1$ SQCD with the superpotential $W={\rm Tr}\,(\,{\ov Q}m_Q Q)$ and in its dual variant \cite{S1, S2}.  It was shown that the direct theory and its Seiberg dual variant are not equivalent in both scenarios. In this paper we calculate the mass spectra in the $\mathbf\Phi$ and $\mathbf{d\Phi}$ theories within the scenario $\#2$. As will be shown below, similarly to the standard SQCD, the use of the small parameter $\bd/N_F\ll 1$ allows to trace explicitly the parametric differences in mass spectra of the direct and dual theories.

\section{Direct theory. Unbroken flavor symmetry}

\hspace*{1cm}{\bf 2.1\,\,\, L - vacua}\\

In these $(2N_c-N_F)$ vacua the current quark mass at $\la\ll\mph\ll\mo$ looks as, see (1.5),
\bq
\langle m^{\rm tot}_Q\rangle_L\equiv\langle m^{\rm tot}_Q(\mu=\la)\rangle_{L}=m_Q-\langle\Phi\rangle_L= m_Q+\frac{\nd}{N_c}\frac{\qql}{\mph}\,,
\eq
\bq
\qql\sim\la^2\Bigl (\frac{\la}{\mph}\Bigr )^{\frac{\nd}{2N_c-N_F}}\gg m_Q\mph,\quad
\langle m^{\rm tot}_Q\rangle_L\sim \la\Bigl (\frac{\la}{\mph}\Bigr )^{\frac{N_c}{2N_c-N_F}}\,,\nonumber
\eq
\bq
\mql=\frac{\langle m^{\rm tot}_Q\rangle_{L}}{z_Q(\la,m^{\rm pole}_{Q,\,L})}\sim \la \Bigl (\frac{\la}{\mph}\Bigr )^{\frac{N_F}{3(2N_c-N_F)}}\sim\lym^{(L)}\,,\quad z_Q(\la,\mu\ll\la)\sim\Bigl (\frac{\mu}{\la}\Bigr )
^{\bo/N_F}\ll 1\,.\nonumber
\eq
We compare $\mql$ with the gluon mass due to possible higgsing of quarks. This last looks as
\bq
\mgl^2\sim \Bigl (a_{*}\sim 1\Bigr ) z_Q(\la,\mgl)\qql \quad \ra \quad \mgl\sim \mql\sim\lym^{(L)}=\langle S\rangle^{1/3}_L\,.
\eq
Hence, qualitatively, the situation is the same as in the standard SQCD \cite{ch2}. And one can use here the same reasonings, see the footnote 3 in \cite{ch2}. In the case considered, there are only $(2N_c-N_F)$ these isolated L vacua with {\it unbroken flavor symmetry}. If quarks were higgsed in these L vacua, then {\it the flavor symmetry will be necessary broken spontaneously} due to the rank restriction because $N_F>N_c$, and there will appear the genuine exactly massless Nambu-Goldstone fields $\Pi$ (pions), so that there will be a continuous family of non-isolated vacua. This is "the standard point of tension" in the dynamical scenario $\#2$, see \cite{ch2}. Therefore, as in \cite{ch2}, to save this scenario $\#2$, here and everywhere below in a similar situations we will assume that $\mg=\mql/(\rm several)$, so that quarks are not higgsed but are in the HQ (heavy quark) phase and are confined. Otherwise, this scenario $\#2$ should be rejected at all.

Therefore (see sections 3, 4 in \cite{ch2}), after integrating out all quarks as heavy ones at $\mu<\mql$ and then all $SU(N_c)$ gluons at $\mu<\lym^{(L)}=\mql/(\rm several)$ via the Veneziano-Yankielowicz (VY) procedure \cite{VY},  we obtain the Lagrangian of fions
\bq
K=z_{\Phi}(\la,\mql){\rm Tr}\,(\Phi^\dagger\Phi)\,, \quad z_{\Phi}(\la,\mql)\sim \frac{1}{z^2_Q(\la,\mql)}\sim\Bigl (\frac{\la}{\mql}\Bigr )^{2\bo/N_F}\gg 1\,,
\eq
\bq
W=N_c S+\frac{\mph}{2}\Biggl [{\rm Tr}\,(\Phi^2)-\frac{1}{\nd}\Bigl ({\rm Tr}\,\Phi\Bigr )^2\Biggr ]\,,
\quad S=\Bigl (\la^{\bo}\det m^{\rm tot}_{Q,L}\Bigr )^{1/N_c}\,,\quad m^{\rm tot}_{Q,L}=(m_Q-\Phi)\,,\nonumber
\eq
and one has to choose the L - vacua in (2.3).

There are two contributions to the mass of fions in (2.3), the perturbative one from the term $\sim \mph\Phi^2$ in $W$
and the non-perturbative one from $\sim S$, and both are parametrically the same, $\sim \lym^{(L)}\gg m_Q$. Therefore,
\bq
\mu(\Phi)\sim\frac{\mph}{z_{\Phi}(\la,\mql)}\sim\mql\sim\lym^{(L)}\,.
\eq
Besides, see (1.14), because in these L - vacua
\bq
\mu^{\rm conf}_o\sim\la \Bigl (\frac{\la}{\mph}\Bigr )^{\frac{N_F}{3(2N_c-N_F)}}\sim\mql\sim\lym^{(L)}\,,
\eq
and fions are dynamically irrelevant at $\mu^{\rm conf}_o<\mu<\la$ and can become relevant only at the scale $\mu<
\mu^{\rm conf}_o$, it remains unclear whether there is a pole in the fion propagators at $p\sim\mu^{\rm conf}_o\sim\mql$.  May be yes but maybe not, see section 4 in \cite{ch2}.\\

On the whole for the mass spectrum in these L - vacua. The quarks ${\ov Q}, Q$ are confined and strongly coupled here, the coupling being $a_*\sim 1$. Parametrically, there is only one scale $\sim \lym^{(L)}$ in the mass spectrum at $\la\ll\mph\ll\mo$. And there is no parametrical guaranty that there is the second generation of fions with
the pole masses $\mu_2^{\rm pole}(\Phi)\sim\lym^{(L)}$.

The condensate $\langle{\ov Q}Q\rangle_L$ and the quark pole mass $\mql$ become frozen at their SQCD values at $\mph\gg
\mo,\, \langle{\ov Q}Q\rangle_{SQCD}\sim\la^2(m_Q/\la)^{\nd/N_c},\,\, m^{\rm pole}_{SQCD}\sim \lym^{(SQCD)}\sim\la(m_Q/
\la)^{N_F/3N_c}$ \cite{ch2}, while $\mph$ increases and $\mu^{\rm conf}_o\ll m^{\rm pole}_{Q,SQCD}$ decreases, see (1.14). Hence, the perturbative contribution $\sim\mph/z_{\Phi}(\la,\mql)\gg m^{\rm pole}_{Q,SQCD}$ to the fion mass becomes dominant at $\mph\gg\mo$ and the fion fields will be dynamically irrelevant at $\mu<\la$.\\

Finally, it is worth noting for what follows that, unlike the dual theory, {\it in all vacua of the direct theory the mass spectra remain parametrically the same at $\,\bd/N_F=O(1)$ or $\,\bd/N_F\ll 1$}.\\

{\bf 2.2\,\,\, S - vacua}\\

In these $\nd$ vacua the quark mass at $\la\ll\mph\ll\mo$ looks as, see (1.7),
\bq
\frac{\langle m^{\rm tot}_Q(\mu=\la)\rangle_S}{\la}\sim \frac{\langle S\rangle_S}{\la\langle{\ov Q}Q\rangle_S}\sim \Bigl (\frac{\langle{\ov Q}Q\rangle_S}{\la^2}\Bigr )^{N_c/\nd}\sim \Bigl (\frac{m_Q\mph}{\la^2}\Bigr )^{N_c/\nd}\,,\nonumber
\eq
\bq
\mqs\sim \la\Bigl (\frac{m_Q\mph}{\la^2}\Bigr )^{N_F/3\nd}\sim\lym^{(S)}=\langle S\rangle^{1/3}_S,\quad \la\ll \mph\ll\mo\,.
\eq
This has to be compared with the gluon mass due to possible higgsing of quarks
\bq
\mgs^2\sim (a_*\sim 1)z_Q(\la,\mgs)\qqs\,\ra\,\mgs\sim\mqs\sim\lym^{(S)},\, z_Q(\la,\mgs)\sim\Bigl (\frac{\mgs}{\la}\Bigr )^{\frac{\bo}{N_F}}.
\eq

For the same reasons as in previous section, it is clear that quarks will not be higgsed in these vacua at $N_F>N_c$ (as otherwise the flavor symmetry will be broken spontaneously). Hence, as in \cite{ch2}, we assume here also that the pole mass of quarks is the largest physical mass, $\mu_H=\mqs=(\rm several)\mgs$.

But, in contrast with the L - vacua, the fion fields {\it become dynamically relevant in these S - vacua} at scales $\mu<\mu^{\rm conf}_o$, see (1.14) and section 4 in \cite{ch1}, if
\bq
\mu^{\rm conf}_o\sim\la\Bigl (\frac{\la}{\mph}\Bigr )^{\frac{N_f}{3(2N_c-N_F)}}\gg\mqs \quad\ra \quad {\rm i.e.\,\, at}\quad  \la\ll\mph\ll\mo\,.
\eq

Therefore, there is the second generation of $N_F^2$ fions with the pole masses
\bq
\mu_2^{\rm pole}(\Phi)\sim \mu_o^{\rm conf}\gg\mqs\sim\lym^{(S)}\,.
\eq

Nevertheless \cite{ch1}, the theory remains in the conformal regime and the quark and fion anomalous dimensions remain the same in the whole range of $\mqs<\mu<\la$ of scales, but fions become effectively massless at $\mu<\mu^{\rm conf}_o$ and begin to contribute to the 't Hooft triangles.

The RG evolution of the quark and fion fields becomes frozen at scales $\mu<\mqs$ because the heavy quarks decouple. Proceeding as before, i.e. integrating out first all quarks as heavy ones at $\mu<\mqs=(\rm several) \lym^{(S)}$ and then all $SU(N_c)$ gluons at $\mu<\lym^{(S)}$, one obtains the Lagrangian of fions as in (2.3), with a replacement $z_Q(\la,\mql)\ra z_Q(\la,\mqs)$ (and the S-vacua have to be chosen therein).

Because fions became relevant at $\mqs\ll\mu\ll\mu^{\rm conf}_o$, one could expect that their running mass will be much smaller than $\mqs$. This is right, but only for $\mu^{\rm pert}_{\Phi}\sim \mph/z_Q(\la,\mqs)\ll\mqs$. But there is also additional {\it non-perturbative contribution} to the fion mass originating from the region of scales $\mu\sim\mqs$ and it is dominant in these S - vacua,
\bq
\mu(\Phi)\sim\frac{1}{z_{\Phi}(\la,\mqs)}\,\frac{\langle S\rangle_S}{\langle m^{\rm tot}_Q\rangle^2_S}\sim\mqs\,,\quad
z_{\Phi}(\la,\mqs)\sim \Bigl (\frac{\la}{\mqs}\Bigr )^{2\bo/N_F}\,.
\eq
Therefore, despite the fact that the fions are definitely dynamically relevant in the range of scales $\mqs\ll\mu\ll\mu^{\rm conf}_o\ll\la$ at $\la\ll\mph\ll\mo$, whether there is the third generation of fions, i.e. whether there is a pole in the fion propagator at $p=\mu_3^{\rm pole}(\Phi)\sim\mqs\sim\lym^{(S)}$ remains unclear. \\

On the whole for the mass spectra in these S - vacua. The largest are the masses of the second generation fions, $\mu_2^{\rm pole}(\Phi)\sim\la\Bigl (\la/\mph\Bigr )^{N_F/3(2N_c-N_F)}\gg\mqs$. The scale of all other masses is $\sim\mqs\sim\lym^{(S)}$, see (2.6). There is no parametrical guaranty that there is the third generation of fions with the pole masses $\mu_3^{\rm pole}(\Phi)\sim\lym^{(S)}$. May be yes, but maybe not.\\

The vacuum condensates $\langle{\ov Q}Q\rangle_{S}$ and $\mqs$ evolve into their independent of $\mph$ SQCD-values at $\mph\gg\mo$,
\bq
\langle{\ov Q}Q\rangle_{SQCD}\sim \la^2\Bigl (\frac{m_Q}{\la}\Bigr )^{\nd/N_c}\,,\quad m^{\rm pole}_{Q,SQCD}\sim\la\Bigl (\frac{m_Q}{\la}\Bigr )^{N_F/3N_c}\,,
\eq
and the perturbative contribution $\sim \mph/z_Q(\la,m^{\rm pole}_{Q,SQCD})$ to the fion mass becomes dominant. Hence, because $m^{\rm pole}_{Q,SQCD}\gg\mu^{\rm conf}_o$, the fions fields become dynamically irrelevant at all scales $\mu<\la$ when $\mph\gg\mo$.\\

\section{Dual theory. Unbroken flavor symmetry}

\hspace*{1cm}{\bf 3.1\,\,\, $\bf L$ - vacua,\,\,\, $\bd/N_F\ll 1$}\\

Because $\la^2/\mph\ll\la$, the mions are effectively massless and dynamically relevant at $\mu\sim\la$ (and so in some range of scales below $\la$). By definition, $\mu\sim\la$ is such a scale that the dual theory already entered sufficiently deep the conformal regime, i.e. the dual gauge coupling ${\ov a}(\mu<\la)
=\nd{\ov\alpha}(\mu<\la)/2\pi$ is sufficiently close to its small frozen value,  $\ov\delta=[{\ov a}_*-{\ov a}(\mu\sim\la)]/{\ov a}_*\ll 1$, and $\ov\delta$ is neglected everywhere below in comparison with 1 (and the same for the Yukawa coupling ${\ov a}_f=\nd{\ov\alpha}_f/2\pi$), see \cite{ch2} and appendix therein. The fixed point value of the dual gauge coupling is ${\ov a}_*\simeq 7\bd/3\nd\ll 1$ \cite{KSV}.

We recall also that the mion condensates are matched to the condensates of direct quarks in all vacua, $\langle M_{ij}(\mu=\la)\rangle=\langle{\ov Q}_j Q_i(\mu=\la)\rangle$\,. Hence, in these L - vacua
\bq
\langle M\rangle_L\sim\la^2\Bigl (\frac{\la}{\mph}\Bigr )^{\frac{\nd}{2N_c-N_F}}\,,\quad
\langle N\rangle_L=\frac{Z_q\la\langle S\rangle_L}{\langle M\rangle_L}\sim Z_q\la^2 \Bigl (\frac{\la}{\mph}\Bigr )^{\frac{N_c}{2N_c-N_F}}\,,
\eq
\bq
Z_q\sim\exp \Bigl\{-\frac{1}{3{\ov a}_{*}}\Bigr\}\sim \exp \Bigl\{-\frac{\nd}{7\bd}\Bigr\}\ll 1\,.\nonumber
\eq
The current mass $\mu_{q,L}\equiv \mu_{q,L}(\mu=\la)$ of dual quarks ${\ov q},\,q$ and their pole mass in these $(2N_c-N_F)$ L - vacua are, see (1.17),(3.1),
\bq
\frac{\mu_{q,L}}{\la}=\frac{\langle M\rangle_L}{Z_q\la^2}\sim\frac{1}{Z_q}\Bigl (\frac{\la}{\mph}\Bigr )^{\frac{\nd}{2N_c-N_F}}\,,\quad
\mu^{\rm pole}_{q,L}=\frac{\mu_{q,L}}{z_q(\la,\mu^{\rm pole}_{q,L})}\,,\quad z_q(\la,\mu^{\rm pole}_{q,L})\sim
\Bigl (\frac{\mu^{\rm pole}_{q,L}}{\la}\Bigr )^{\bd/N_F}\,,\nonumber
\eq
\bq
\mu^{\rm pole}_{q,L}\sim\la\Bigl (\frac{\mu_{q,L}}{\la}\Bigr )^{N_F/3\nd}\sim\frac{\la}{Z_q}\Bigl (\frac{\la}{\mph}\Bigr )^{\frac{N_F}{3(2N_c-N_F)}}\sim\frac{1}{Z_q}\lym^{(L)}\gg\lym^{(L)} \,, \quad \la\ll\mph\ll\mo\,,
\eq
\bq
\frac{\mu_{q,L}}{\la}\sim\frac{1}{Z_q}\Bigl (\frac{m_Q}{\la}\Bigr )^{\frac{\nd}{N_c}}\,,\quad \mu^{\rm pole}_{q,L}\sim\frac{\la}{Z_q}\Bigl (\frac{m_Q}{\la}\Bigr )^{\frac{N_F}{3N_c}}\sim\frac{1}{Z_q}\lym^{(SQCD)}
\gg \lym^{(SQCD)}\,,\quad \mph\gg\mo\,,\nonumber
\eq
while the gluon mass due to possible higgsing of dual quarks looks at $\la\ll\mph\ll\mo$ as
\bq
{\ov\mu}_{\rm gl, L}\sim\Bigl [{\ov a}_*\langle N\rangle_{L}\, z_q(\la,\mug)\Bigr]^{1/2}\sim Z_q^{1/2}\la\Bigl (\frac{\la}{\mph}\Bigr )^{N_F/3(2N_c-N_F)}\sim Z_q^{3/2}\mu^{\rm pole}_{q,L}\ll \mu^{\rm pole}_{q,L}\,.
\eq
Therefore, the dual quarks are definitely in the HQ phase in these L - vacua at $\bd/\nd\ll 1$.

With decreasing scale the perturbative current mass $\mu_{M}(\mu)$ of mions
\bq
\mu_{M}(\mu)\sim\frac{Z_q^2\la^2}{\mph z_M(\mu)}=\frac{Z_q^2\la^2}{\mph}\Bigl (\frac{\mu}{\la}\Bigr )^{2\bd/N_F}
\eq
decreases but more slowly than the scale $\mu$ itself, because $\gamma_M= - (2{\rm\bd}/N_F),\,\,|\gamma_M|<1$ at $3/2<N_F/N_c<2$\,, and becomes frozen at $\mu<\mu^{\rm pole}_{q,L}$,\, $\mu_M(\mu<\mu^{\rm pole}_{q,L})=\mu_M(\mu=\mu^{\rm pole}_{q,L})$.

After integrating out all dual quarks as heavy ones at $\mu<\mu^{\rm pole}_{q,L}$ and then all $SU(\nd)$ gluons at $\mu<\lym^{(L)}$ via the Veneziano-Yankielowicz (VY) procedure \cite{VY}, the Lagrangian of mions looks as
\bq
K=\frac{z^{(L)}_M(\la,\mu^{\rm pole}_{q,L})}{Z_q^2\la^2}{\rm Tr}\,(M^\dagger M)\,,\quad z^{(L)}_M(\la,\mu^{\rm pole}_{q,L})\sim\Bigl (\frac{\la}{\mu^{\rm pole}_{q,L}}\Bigr )^{2\bd/N_F}\,,\quad
S=\Biggl (\,\frac{\det M}{\la^{\bo}}\,\Biggr )^{1/\nd},\nonumber
\eq
\bq
W= -\nd S+m_Q{\rm Tr}\,M -\frac{1}{2\mph}\Biggl [{\rm Tr}\, (M^2)- \frac{1}{N_c}({\rm Tr}\, M)^2  \Biggr ]\,.
\eq

There are two contributions to the mass of mions in (3.5), the perturbative one from the term $\sim M^2/\mph$ in $W$
and non-perturbative one from $\sim S$. Both are parametrically the same and the total contribution looks as
\bq
\mu^{\rm pole}(M)\sim \frac{Z_q^2\la^2}{z_M(\la,\mu^{\rm pole}_{q,L})\mph}\sim Z_q^2\lym^{(L)}\ll \lym^{(L)}\ll \mu^{\rm pole}_{q,L}\,,
\eq
and this parametrical hierarchy guarantees that the mass $\mu^{\rm pole}(M)$ in (3.6) is indeed the {\it pole mass} of mions.\\

On the whole, the mass spectrum in these dual L - vacua looks as follows at $\la\ll\mph\ll\mo$. a) There is a large number of heaviest flavored hadrons made of weakly interacting and weakly confined (the string tension being $\sqrt\sigma
\sim\lym^{(L)}\ll\mu^{\rm pole}_{q,L}$) non-relativistic quarks ${\ov q}, q$ with the pole masses $\mu^{\rm pole}_{q,L}/\lym^{(L)}\sim \exp (\nd/7\bd)\gg 1$. The mass spectrum of low-lying flavored mesons is Coulomb-like with parametrically small mass differences $\Delta\mu_H/\mu_H=O(\bd^2/\nd^2)\ll 1$. b) A large number of gluonia made of $SU(\nd)$ gluons with the mass scale $\sim\lym^{(L)}\sim\la (\la/\mph)^{N_F/3(2N_c-N_F)}$. c) $N_F^2$ lightest mions with parametrically smaller masses $\mu^{\rm pole}(M)/\lym^{(L)}\sim \exp (-2\nd/7\bd)\ll 1$.

At $\mph\gg\mo$ these L - vacua evolve into the vacua of the dual SQCD theory (dSQCD), see section 4 in \cite{ch2}.\\
\newpage

{\bf 3.2\,\,\, $\bf S$ - vacua,\,\,\, $\bd/N_F\ll 1$}\\

The current mass $\mu_{q,S}\equiv \mu_{q,S}(\mu=\la)$ of dual quarks ${\ov q},\,q$ at the scale $\mu=\la$ in these $(N_F-N_c)$ dual S-vacua is, see (1.17),(1.7),
\bq
\mu_{q,S}=\frac{\langle M\rangle_S}{Z_q\la}\sim\frac{m_Q\mph}{Z_q\la}\,,\quad Z_q\sim \exp \Bigl\{-\frac{\nd}{7\bd}\Bigr\}\ll 1\,.
\eq

In comparison with the L - vacua in section 3.1, a qualitatively new element here is that $\mu^{\rm pole}(M)$ is the largest mass, $\mu^{\rm pole}(M)\gg\mu^{\rm pole}_{q,S}$, in the wide region $\la\ll\mph\ll Z_q^{\,3/2}\mo$ (see (3.14) below). In this region: a) the mions are effectively massless and dynamically relevant at scales $\mu^{\rm pole}(M)\ll\mu\ll\la$,\,\, b) there is a pole in the mion propagator at the momentum $p=\mu^{\rm pole}(M)$,
\bq
\mu^{\rm pole}(M)=\frac{Z_q^2\la^2}{z_M(\la,\mu^{\rm pole}(M))\mph}\,,\quad z_M(\la,\mu^{\rm pole}(M))\sim\Bigl (\frac{
\la}{\mu^{\rm pole}(M)}\Bigr )^{2\bd/N_F}\,,\nonumber
\eq
\bq
\mu^{\rm pole}(M)\sim Z_q^2\la\Bigl (\frac{\la}{\mph}\Bigr )^{\frac{N_F}{3(2N_c-N_F)}}\,,\quad Z_q^2\sim \exp\{-\frac{2\nd}{7\bd}\}\ll 1\,.
\eq

The mions then become too heavy and dynamically irrelevant at $\mu\ll\mu^{\rm pole}(M)$. Due to this, they decouple from the RG evolution of dual quarks and gluons and from the 't Hooft triangles, and (at $\mph$ not too close to $\mo$ to have enough "time" to evolve, see (3.13)\,) the remained dual theory of $N_F$ quarks ${\ov q}, q$ and $SU(\nd)$ gluons evolves into a new conformal regime with a new smaller value of the frozen gauge coupling, ${\ov a}^{\,\prime}_*\simeq
\bd/3\nd={\ov a}_*/7\ll 1$. It is worth noting that, in spite of that  mions are dynamically irrelevant at $\mu<\mu^{\rm pole}(M)$, their renormalization factor $z_M(\mu<\mu^{\rm pole}(M))$ still runs in the range of scales $\mu^{\rm pole}_{q,S}<\mu<\mu^{\rm pole}(M)$ being induced by loops of still effectively massless quarks and gluons.

The next physical scale is the perturbative pole mass of dual quarks
\bq
\mu^{\rm pole}_{q,S}=\frac{\langle M\rangle_S}{Z_q\la}\,\frac{1}{z_q(\la,\mu^{\rm pole}_{q,S})}\,,\quad
z_q(\la,\mu^{\rm pole}_{q,S})=\Bigl (\frac{\mu^{\rm pole}_{q,S}}{\la}\Bigr )^{\bd/N_F}{\ov\rho}_S\,,\nonumber
\eq
\bq
{\ov\rho}_S=\Bigl (\frac{{\ov a}_*}{{\ov a}^{\,\prime}_*}\Bigr )^{\frac{\nd}{N_F}}\exp\Bigl\{\frac{\nd}{N_F}
\Bigl (\frac{1}{{\ov a}_*}-\frac{1}{{\ov a}^{\,\prime}_*}\Bigr )\Bigr\}\sim \frac{{\ov Z}_q}{Z_q}\ll 1\,,\quad
{\ov Z}_q\sim \exp \Bigl\{-\frac{\nd}{\bd}\Bigr\}\sim \Bigl (Z_q\Bigr )^{7}\ll Z_q\,,\nonumber
\eq
\bq
\mu^{\rm pole}_{q,S}\sim \frac{1}{{\ov Z}_q}\lym^{(S)}\gg\lym^{(S)}\,,\quad \lym^{(S)}=\la\Bigl (\frac{m_Q\mph}{\la^2}
\Bigr )^{N_F/3\nd}\,.
\eq

This has to be compared with the gluon mass due to possible higgsing of ${\ov q}, q$
\bq
{\ov\mu}^{\,2}_{\rm gl, S}\sim \langle N\rangle_S\, z_q(\la,{\ov\mu}_{\rm gl, S})\,\,\ra\,\,{\ov\mu}_{\rm gl, S}\sim
{\ov Z}_q^{\,1/2}\lym^{(S)}\ll\lym^{(S)}\ll\mu^{\rm pole}_{q,S}.
\eq
The parametrical hierarchy in (3.10) guarantees that the dual quarks are in the HQ phase in these S - vacua.

Hence, after integrating out all quarks at $\mu<\mu^{\rm pole}_{q,S}$ and, finally, $SU(\nd)$ gluons at $\mu<\lym^{(S)}$,
the Lagrangian looks as in (3.5) but with a replacement $z^{(L)}_M (\la,\mu^{\rm pole}_{q,L})\ra z^{(S)}_M (\la,\mu^
{\rm pole}_{q,S})$,
\bq
z^{(S)}_M(\la,\mu^{\rm pole}_{q,S})=\frac{a_f(\mu=\la)}{a_f(\mu=\mu^{\rm pole}_{q,S})}\frac{1}{z^2_q(\la,\mu^{\rm pole}_{q,S})}\sim\frac{1}{z^2_q(\la,\mu^{\rm pole}_{q,S})}\sim
\frac{Z_q^{\,2}}{{\ov Z}_q^{\,2}}\Bigl (\frac{\la}{\mu^{\rm pole}_{q,S}}\Bigr )^{2\bd/N_F}\,.
\eq
The contribution of the term $\sim M^2/\mph$ in the superpotential (3.5) to the frozen low energy value $\mu(M)$
of the running mion mass is dominant at $\mph/\mo\ll 1$ and is
\bq
\hspace*{-5mm}\mu(M)=\frac{Z_q^2\la^2}{z^{(S)}_M(\la,\mu^{\rm pole}_{q,S})\mph}\sim
\frac{{\ov Z}_q^{\,2}\la^2}{\mph}\Bigl (\frac{\mu^{\rm pole}_{q,S}}{\la}\Bigr )^{\frac{2\bd}{N_F}}\ll\mu^{\rm pole}(M)\,.
\eq
The requirement of self-consistency looks in this case as
\bq
\frac{\mu(M)}{\mu^{\rm pole}_{q,S}}\sim {\ov Z}_q^{\,3}\Bigl (\frac{\mo}{\mph}\Bigr )^{N_c/\nd}\gg 1\quad\ra\quad
\frac{\mph}{\mo}\ll {\ov Z}^{\, 3/2}_q\sim\exp \Bigl\{-\frac{3\nd}{2\bd}\Bigr\}\ll Z_q^{3/2}\,,
\eq
the meaning of (3.13) is that only at this condition the range of scales between $\mu^{\rm pole}(M)$ in (3.8) and $\mu^{\rm pole}_{q,S}\ll\mu^{\rm pole}(M)$ in (3.9) is sufficiently large that theory has enough "time" to evolve from ${\ov a}_*=7\bd/3\nd$ to ${\ov a}^{\,\prime}_*=\bd/3\nd$. There is no pole in the mion propagator at the momentum $p=\mu(M)
\gg\mu^{\rm pole}_{q,S}$.

The opposite case with $\mu^{\rm pole}_{q,S}\gg\mu^{\rm pole}(M)$ is realized if the ratio $\mph/\mo$ is still $\ll 1$
but is much larger than $Z_q^{\,3/2}\gg{\ov Z}_q^{\,3/2}$, see (3.14) below. In this case the theory at $\mu^{\rm pole}_{q,S}<\mu<\la$ remains in the conformal regime with ${\ov a}_*=7\bd/3\nd$ and the largest mass is $\mu^{\rm pole}_{q,S}$. One has in this case instead of (3.8),(3.9),(3.13)
\bq
{\ov\rho}_S\sim 1\,,\quad \mu^{\rm pole}_{q,S}\sim\frac{1}{Z_q}\lym^{(S)}\,,\quad
\frac{\mu^{\rm pole}(M)}{\mu^{\rm pole}_{q,S}}\sim Z^3_q\Bigl (\frac{\mo}{\mph}\Bigr )^{N_c/\nd}\,,
\eq
\bq
\frac{\mu^{\rm pole}(M)}{\mu^{\rm pole}_{q,S}}\ll 1\quad\ra\quad Z_q^{\,3/2}\ll\frac{\mph}{\mo}\ll 1\,.\nonumber
\eq

On the whole, the mass spectrum in these $\nd$ dual S - vacua looks as follows at $\la\ll\mph\ll{\ov Z}_q^{\,3/2}\mo$. a) The heaviest are $N_F^2$ mions with the pole masses (3.8). b) There is a large number of flavored hadrons made of weakly interacting and weakly confined (the string tension being $\sqrt\sigma\sim\lym^{(S)}\ll\mu^{\rm pole}_{q,S}$) non-relativistic dual quarks ${\ov q}, q$ with the perturbative pole masses (3.9). The mass spectrum of low-lying flavored mesons is Coulomb-like with parametrically small mass differences $\Delta\mu_H/\mu_H=O(\bd^2/\nd^2)\ll 1$. b) A large number of gluonia made of $SU(\nd)$ gluons with the mass scale $\sim\lym^{(S)}\sim\la (m_Q\mph/\la^2)^{N_F/3\nd}$.

The mions with the pole masses (3.8) remain the heaviest ones, $\mu^{\rm pole}(M)\gg\mu^{\rm pole}_{q,S}$, at values $\mph$ in the range ${\ov Z}_q^{\,3/2}\mo\ll\mph\ll Z_q^{\,3/2}\mo$, while the value $\mu^{\rm pole}_{q,S}$ varies in a range $\lym^{(S)}/Z_q\ll\mu^{\rm pole}_{q,S}\ll\lym^{(S)}/{\ov Z}_q$\,. Finally, in a close vicinity of $\mo,\,\, Z_q^{\,3/2}\mo\ll\mph\ll\mo$, the perturbative pole mass of quarks, $\mu^{\rm pole}_{q,S}\sim\lym^{(S)}/Z_q\gg\lym^{(S)}$, becomes the largest one, while the pole masses of mions $\mu^{\rm pole}(M)\ll\lym^{(S)}$ become as in (3.14).

At $\mph\gg\mo$ these S - vacua evolve into the vacua of dSQCD, see section 4 in \cite{ch2}.

\section{Direct theory. Broken flavor symmetry.\\{\hspace*{1.2cm}} The region $\mathbf{\la\ll\mph\ll\mo}$ }

\hspace*{1cm}{\bf 4.1 \,\,\,  L - type} vacua\\

The quark condensates are parametrically the same as in the L - vacua with unbroken flavor symmetry in section 3.1,
\bq
(1-\frac{n_1}{N_c})\langle\qqo\rangle_{Lt}\simeq -(1-\frac{n_2}{N_c})\langle\qqt\rangle_{Lt},\quad \langle S\rangle=
\frac{\langle\qqo\rangle\langle\qqt\rangle}{\mph}\,,
\eq
\bq
\langle\qqo\rangle_{Lt}\sim\langle\qqt\rangle_{Lt}\sim\la^2\Bigl (\frac{\la}{\mph}\Bigr )^{\frac{\nd}{2N_c-N_F}}\,.\nonumber
\eq

All quarks are in the HQ  phase and are confined and the Lagrangian of fions looks as in (2.3), but one has to choose
the L - type vacua with the broken flavor symmetry in (2.3). Due to this, see (1.5), the masses of hybrid fions
$\Phi_{12}, \Phi_{21}$ are qualitatively different, they are the Nambu-Goldstone particles here and are massless. The "masses" of $\Phi_{11}$ and $\Phi_{22}$ are parametrically as in (2.4),
\bq
\mu(\Phi_{11})\sim\mu(\Phi_{22})\sim\frac{\mph}{z_{\Phi}(\la, m^{\rm pole}_Q)}\sim m^{\rm pole}_{Q,1}\sim m^{\rm pole}_{Q,2}\sim\lym^{(L)}\sim\la \Bigl (\frac{\la}{\mph}\Bigr )^{\frac{N_F}{3(2N_c-N_F)}}\,,
\eq
and hence there is no guaranty that these are the pole masses of fions, see section 4 in \cite{ch1}. May be yes, but maybe not.

On the whole, there are only two characteristic scales in the mass spectra in these L - type vacua. The hybrid fions $\Phi_{12}, \Phi_{21}$ are massless while all other masses are $\sim\lym^{(L)}$.\\

\hspace*{1cm}{\bf 4.2 \,\,\,   $\rm\bf br2$} vacua\\

The condensates of quarks look as
\bq
\langle\qqt\rangle_{\rm br2}\simeq \Bigl (\rho_2=-\frac{n_2-N_c}{N_c}\Bigr )m_Q\mph,\,\,\, \langle\qqo\rangle_{\rm br2}
\sim \la^2\Bigl(\frac{\mph}{\la}\Bigr )^{\frac{n_2}{n_2-N_c}}\Bigl (\frac{m_Q}{\la}\Bigr )^{\frac{N_c-n_1}{n_2-N_c}},
\eq
\bq
\frac{\langle\qqo\rangle_{\rm br2}}{\langle\qqt\rangle_{\rm br2}}\sim \Bigl (\frac{\mph}{\mo}\Bigr)^{\frac{N_c}
{n_2-N_c}}\ll 1\nonumber
\eq
in these vacua with $n_2>N_c\,, n_1<\nd$\,. Hence, the largest among the masses smaller than $\la$ are the masses of the $N_F^2$ second generation fions, see (1.14),
\bq
\mu^{\rm pole}_2(\Phi_{ij})=\mu_o^{\rm conf}\sim \la\Bigl (\frac{\la}{\mph}\Bigr )^{\frac{N_F}{3(2N_c-N_F)}}\,,
\eq
while some other possible characteristic masses look here as
\bq
\langle m^{\rm tot}_{Q,2}\rangle_{\rm br2}=\frac{\langle\qqt \rangle_{\rm br2}}{\mph}\sim m_Q\,,\quad m^{\rm pole}_{Q,1}\sim\la\Bigl(\frac{m_Q}{\la}\Bigr )^{N_F/3N_c}\gg {\tilde m}^{\rm pole}_{Q,2}\,,
\eq
\bq
\mgt^2\sim z_Q(\la,\mgt)\langle\qqt\rangle_{\rm br2},\quad z_Q(\la,\mgt)\sim\Bigl (\frac{\la}{\mgt}\Bigr )^{\frac{\bo}{N_F}}\ll 1\,,
\nonumber
\eq
\bq
\mgt\sim\la\Bigl(\frac{m_Q\mph}{\la^2}\Bigr )^{N_F/3\nd}\gg\mgo\,,\quad \frac{\mgt}{m^{\rm pole}_{Q,1}}\sim
\Bigl (\frac{\mph}{\mo}\Bigr )^{\frac{N_F}{3\nd}}\ll 1\,,
\eq
where $m^{\rm pole}_{Q,1}$ and ${\tilde m}^{\rm pole}_{Q,2}$ are the pole masses of quarks ${\ov Q}_1, Q_1$ and ${\ov Q}_2, Q_2$ and $\mgo,\, \mgt$ are the gluon masses due to possible higgsing of these quarks. Hence, the largest mass is $m^{\rm pole}_{Q,1}$ and the overall phase is $HQ_1-HQ_2$.

The lower energy theory at $\mu<m^{\rm pole}_{Q,1}$ has $N_c$ colors and $N_F^\prime =n_2>N_c$ flavors of quarks ${\ov Q}_2, Q_2$. In the range of scales $m^{\rm pole}_{Q,2}<\mu<m^{\rm pole}_{Q,1}$, it will remain in the conformal regime at $2n_1<\bd$, while it will be in the strong coupling regime at $2n_1>\bd$, with the gauge coupling $a(\mu\ll m^{\rm pole}_{Q,1})\gg 1$. We do not consider the strong coupling regime in this paper and for this reason we take $\bd/\nd=O(1)$ in this subsection.

After the heaviest quarks ${\ov Q}_1, Q_1$ decouple at $\mu<m^{\rm pole}_{Q,1}$, the pole mass of quarks ${\ov Q}_2, Q_2$ in the lower energy theory looks as
\bq
m^{\rm pole}_{Q,2}=\frac{1}{z^{\,\prime}_Q(m^{\rm pole}_{Q,1},m^{\rm pole}_{Q,2})}\Biggl(\,\frac{\langle\qqo\rangle_{\rm br2}}{\langle\qqt\rangle_{\rm br2}}\, m^{\rm pole}_{Q,1}\,\Biggr )\sim\lym^{(\rm br2)}\,,
\eq
\bq
z^{\,\prime}_Q(m^{\rm pole}_{Q,1},m^{\rm pole}_{Q,2})\sim\Bigl (\frac{m^{\rm pole}_{Q,2}}{m^{\rm pole}_{Q,1}}\Bigr )^{\frac{3N_c-n_2}{N_F}}\ll 1\,.\nonumber
\eq

Hence, after integrating out  quarks ${\ov Q}_1, Q_1$ at $\mu<m^{\rm pole}_{Q,1}$ and then  quarks ${\ov Q}_2, Q_2$ and
$SU(N_c)$ gluons at $\mu<\lym^{(\rm br2)}$, the Lagrangian of fions looks as
\bq
K=z_{\Phi}(\la,m^{\rm pole}_{Q,1})\,{\rm Tr}\,\Bigl [\,\Phi_{11}^\dagger \Phi_{11}+\Phi_{12}^\dagger \Phi_{12}+\Phi_{21}^\dagger \Phi_{21}+z^{\,\prime}_{\Phi}(m^{\rm pole}_{Q,1},m^{\rm pole}_{Q,2})\Phi_{22}^\dagger \Phi_{22}\,\Bigr ]\,,
\eq
\bq
z_{\Phi}(\la,m^{\rm pole}_{Q,1})\sim\Bigl (\frac{\la}{m^{\rm pole}_{Q,1}}\Bigr )^{\frac{2(3N_c-N_F)}
{N_F}}\gg 1\,,\quad
z^{\,\prime}_{\Phi}(m^{\rm pole}_{Q,1},m^{\rm pole}_{Q,2})\sim\Bigl (\frac{m^{\rm pole}_{Q,1}}{m^{\rm pole}_{Q,2}}\Bigr )^{\frac{2(3N_c-n_2)}{N_F}}\gg 1\,,\nonumber
\eq
\bq
W=N_c S+W_{\Phi}\,,\quad m^{\rm tot}_Q= (m_Q-\Phi)\,,\quad
\eq
\bq
S=\Bigl (\la^{\bo}\det m^{\rm tot}_Q\Bigr )^{1/N_c}\,,\quad W_{\Phi}=\frac{\mph}{2}\Bigl ({\rm Tr}\,(\Phi^2)-
\frac{1}{\nd}({\rm Tr}\,\Phi)^2 \,\Bigr ).\nonumber
\eq
From (4.8),(4.9), the main contribution to the mass of the third generation fions $\Phi_{11}$ gives the term $\sim\mph\Phi^2_{11}$,
\bq
\mu^{\rm pole}_3(\Phi_{11})\sim\frac{\mph}{z_{\Phi}(\la,m^{\rm pole}_{Q,1})}\sim\Bigl (\frac{\mph}{\mo}\Bigr )m^{\rm pole}_{Q,1}\,,
\eq
while the third generation hybrid fions $\Phi_{12}, \Phi_{21}$ are massless, $\mu^{\rm pole}_3(\Phi_{12})=\mu^{\rm pole}_3(\Phi_{21})=0$. As for the third generation fions $\Phi_{22}$, the main contribution to their masses comes from
the non-perturbative term $\sim S$ in the superpotential 
\bq
\mu_3(\Phi_{22})\sim\frac{\langle S\rangle}{\langle m^{\rm tot}_{Q,2}\rangle^2}\frac{1}{z_{\Phi}(\la,m^{\rm pole}_{Q,1})
z^{\,\prime}_{\Phi}(m^{\rm pole}_{Q,1},m^{\rm pole}_{Q,2})}\sim m^{\rm pole}_{Q,2}\sim\lym^{(\rm br2)}.
\eq
In such a situation there is no guaranty that there is a pole in the propagator of $\Phi_{22}$ at the momentum $p\sim
m^{\rm pole}_{Q,2}$. May be yes but maybe not, see section 4 in \cite{ch1}.\\

\hspace*{1cm}{\bf 4.3 \,\,\,   \bf Special} vacua,\,\, $n_1=\nd,\,\, n_2=N_c$\\

Nothing specific for the dynamical scenario $\# 1$ with the diquark condensate (DC) was used in a description of these vacua in section 7.3 of \cite{ch1}. Hence, all remains the same in the dynamical scenario $\# 2$ here.\\

\section{Dual theory. Broken flavor symmetry.\\{\hspace*{1.2cm}} The region $\mathbf{\la\ll\mph\ll\mo}$ }

\hspace*{1cm}{\bf 5.1 \,\,\,   \bf L - type} vacua, $\,\,\bd/N_F\ll 1$\\

The condensates of mions and dual quarks look here as
\bq
\langle M_1+M_2-\frac{1}{N_c}{\rm Tr}\,M\rangle_{Lt}=m_Q\mph\quad\ra\quad \frac{\langle M_1\rangle_{Lt}}{\langle M_2\rangle_{Lt}}\simeq -\,\frac{N_c-n_1}{N_c-n_2}\,,\nonumber
\eq
\bq
\langle M_1\rangle_{Lt}\langle N_1\rangle_{Lt}=\langle M_2\rangle_{Lt}\langle N_2\rangle_{Lt}=Z_q\la\langle S
\rangle_{Lt},\quad
\langle S\rangle_{Lt}=\frac{\langle M_1\rangle_{Lt}\langle M_2\rangle_{Lt}}{\mph}\,.\nonumber
\eq

I.e., all condensates are parametrically the same as in the L - vacua with unbroken flavor symmetry in section 3.1 and
the overall phase is also $HQ_1-HQ_2$. The pole masses of dual quarks are as in (3.2), the Lagrangian of mions is
as in (3.5) and the pole masses of mions $M_{11}$ and $M_{22}$ are as in (3.6). But the masses of hybrid mions $M_{12}$
and $M_{21}$ are qualitatively different here. They are the Nambu-Goldstone particles now and are exactly massless,
$\mu(M_{12})=\mu (M_{21})=0$.\\

\hspace*{1cm}{\bf 5.2 \,\,\,   $\rm\bf br2$ vacua}, $\,\,\bd/N_F=O(1)$\\

In these vacua with $n_2>N_c\,, n_1<\nd$ the condensates of mions and dual quarks look as
\bq
\langle M_2\rangle_{\rm br2}\simeq -\,\frac{n_2-N_c}{N_c}\, m_Q\mph\,,\quad \langle M_1\rangle_{\rm br2}
\sim \la^2\Bigl(\frac{\mph}{\la}\Bigr )^{\frac{n_2}{n_2-N_c}}\Bigl (\frac{m_Q}{\la}\Bigr )^{\frac{N_c-n_1}{n_2-N_c}},\,\,
\eq
\bq
\frac{\langle M_1\rangle_{\rm br2}}{\langle M_2\rangle_{\rm br2}}\sim \Bigl (\frac{\mph}{\mo}\Bigr )^{\frac{N_c}{n_2-N_c}}\ll 1\,,\nonumber
\eq
\bq
\langle N_1\rangle_{\rm br2}=\langle{\ov q}_1 q_1(\mu=\la)\rangle_{\rm br2}=\frac{\la\langle S\rangle_{\rm br2}}{\langle M_1\rangle_{\rm br2}}=\frac{\la\langle M_2\rangle_{\rm br2}}{\mph}\sim m_Q\la\gg\langle N_2\rangle_{\rm br2}\,.\nonumber
\eq

From these, the heaviest are $N_F^2$ mions $M_{ij}$ with the pole masses
\bq
\hspace*{-4mm}\mu^{\rm pole}(M)=\frac{\la^2/\mph}{z_M(\la,\mu^{\rm pole}(M))}\sim \la\Bigl (\frac{\la}{\mph}\Bigr )^{\frac{N_F}{3(2N_c-N_F)}},\, z_M(\la,\mu^{\rm pole}(M))\sim\Bigl (\frac{\la}{\mu^{\rm pole}(M)}\Bigr )^{\frac{2\bd}{N_F}}\gg 1\,\,
\eq
while some other possible characteristic masses look as
\bq
\mu_{q,2}=\frac{\langle M_2\rangle}{\la}\sim\frac{m_Q\mph}{\la},\quad {\tilde\mu}^{\rm pole}_{q,2}\sim\la
\Bigl (\frac{m_Q\mph}{\la^2}\Bigr )^{N_F/3\nd}\gg\mu^{\rm pole}_{q,1}\,,
\eq
\bq
{\ov\mu}_{\rm gl,1}\sim\la\Bigl (\frac{\langle N_1\rangle}{\la^2}\Bigr )^{N_F/3N_c}\sim \la\Bigl(\frac{m_Q}{\la}\Bigr )^{N_F/3N_c}\gg{\ov\mu}_{\rm gl,2}\,,\quad\frac{{\ov\mu}_{\rm gl,1}}{{\tilde\mu}^{\rm pole}_{q,2}}
\sim \Bigl(\frac{\mo}{\mph}\Bigr )^{N_F/3\nd}\gg 1\,,\nonumber
\eq
where $\mu^{\rm pole}_{q,1}$ and ${\tilde\mu}^{\rm pole}_{q,2}$ are the pole masses of quarks ${\ov q}_1, q_1$ and ${\ov q}_2, q_2$ and ${\ov\mu}_{\rm gl,1},\, {\ov\mu}_{\rm gl,2}$ are the gluon masses due to possible higgsing of these quarks. Hence, the largest mass is ${\ov\mu}_{\rm gl,1}$ and the overall phase is $Higgs_1-HQ_2$.

After integrating out all higgsed gluons and quarks ${\ov q}_1, q_1$, we write the dual Lagrangian at $\mu=\muo$ as
\bq
K= z_M(\la,\muo){\rm Tr}\,\frac{M^\dagger M}{\la^2}+ z_q(\la,\muo){\rm Tr}\,\Bigl [\,2\sqrt{N_{11}^\dagger N_{11}}+K_{\rm hybr}+\Bigl ({\dq}^{\dagger}_2 {\dq}_2 +({\dq}_2\ra {\odq}_2 )\Bigr )\,\Bigr ]\,,\nonumber
\eq
\bq
K_{\rm hybr}=\Biggl (N^{\dagger}_{12}\frac{1}{\sqrt{N_{11} N^{\dagger}_{11}}} N_{12}+
N_{21}\frac{1}{\sqrt{N^{\dagger}_{11} N_{11}}} N^\dagger_{21}\Biggr ),\quad z_q(\la,\muo)=\Bigl (\frac{\muo}{\la}\Bigr )^{\bd/N_F}\,,
\eq
\bq
z_M(\la,\muo)=1/z^2_q(\la,\muo),\quad W=\Bigl [-\frac{2\pi}{{\ov\alpha}(\mu)}{\ov{\textsf s}}
\Bigr ]-\frac{1}{\la}{\rm Tr}\,\Bigl ({\odq}_2 M_{22}\dq_2\Bigr )- W_{MN}+W_{M},\nonumber
\eq
\bq
W_{MN}=\frac{1}{\la}{\rm Tr}\,\Bigl (M_{11}N_{11}+M_{21} N_{12}+N_{21}
M_{12}+M_{22} N_{21}\frac{1}{N_{11}} N_{12}\Bigr )\,,\nonumber
\eq
where $\odq_2, \dq_2$ are the quarks ${\ov q}_2, q_2$ with unhiggsed colors, $\ov{\textsf s}$ is the field strength of unhiggsed dual gluons and the hybrid nions $N_{12}$ and $N_{21}$ are, in essence, the quarks ${\ov q}_2, q_2$ with higgsed colors, $W_M$ is given in (1.17). The lower energy theory at $\mu<\muo$ has $\nd^{\,\prime}=\nd-n_1$ colors and $n_2>N_c$ flavors, $\bd^{\,\prime}=\bd-2n_1<\bd$. We consider here only the case $\bd^{\,\prime}>0$ when it remains in the conformal window. In this case the value of the pole mass $\qtp$ in this lower energy theory is
\bq
\qtp\sim \frac{\langle M_2\rangle}{\la}\frac{1}{z_q(\la,\muo) z^{\,\prime}_q(\muo,\qtp)}\sim\lym^{(\rm br2)}\,, \quad z^{\,\prime}_q(\muo,\qtp)\sim\Bigl (\frac{\qtp}{\muo}\Bigr )^{\bd^{\,\prime}/n_2}\ll 1\,.
\eq

The fields $N_{11}, N_{12}, N_{21}$ and $M_{11}, M_{12}, M_{21}$ are frozen and do not evolve at $\mu<\muo$.  After integrating out remained unhiggsed quarks $\odq_2, \dq_2$ as heavy ones and unhiggsed gluons at $\mu<\lym^{(\rm br2)}$
the Lagrangian of mions and nions looks as, see (5.4),
\bq
K=z_M(\la,\muo){\rm Tr}\,K_M+z_q(\la,\muo)\Bigl [\,2\sqrt{N_{11}^\dagger N_{11}}+K_{\rm hybr}\,\Bigr ],\,\, z^{\,\prime}_M (\muo,\qtp)\sim\Bigl (\frac{\muo}{\qtp}\Bigr )^{\frac{2\bd^{\,\prime}}{n_2}}\gg 1,\nonumber
\eq
\bq
K_M=\frac{1}{\la^2}\Bigl (M_{11}^\dagger M_{11}+M_{12}^\dagger M_{12}+M_{21}^\dagger M_{21}+
z^{\,\prime}_M (\muo,\qtp)M_{22}^\dagger M_{22}\Bigr )\,,
\eq
\bq
W=-\nd^{\,\prime}S-W_{MN}+W_M\,,\quad S=\Bigl (\lym^{(\rm br2)}\Bigr )^3 \Biggl (\det\frac{\langle N_1\rangle}{N_{11}}
\det\frac{M_{22}}{\langle M_2\rangle}\Biggr )^{1/\nd^{\,\prime}},\quad \lym^{(\rm br2)}\sim
\Bigl (m_Q\langle M_1\rangle\Bigr )^{1/3}.\nonumber
\eq

From (5.6), the "masses" of mions look as
\bq
\mu(M_{11})\sim\mu(M_{12})\sim\mu(M_{21})\sim\frac{\la^2}{z_M(\la,\muo)\mph}\sim\Bigl (\frac{\mo}{\mph}\Bigr )
\muo\gg\muo\,,
\eq
\bq
\mu(M_{22})\sim\frac{\la^2}{z_M(\la,\muo)z^\prime_M (\muo,\qtp)\mph}\sim \Bigl (\frac{\mo}{\mph}\Bigr )^{\frac{3N_c-n_2}{3(n_2-N_c)}}\,\muo\gg \muo\,,
\eq
while the pole masses of nions $N_{11}$ are
\bq
\mu^{\rm pole}(N_{11})\sim \frac{\mph\langle N_1\rangle_{\rm br2}}{z_q(\la,\muo)\la^2}\sim\Bigl (\frac{\mph}{\mo}\Bigr )\muo\,,
\eq
and the hybrid nions $N_{12}, N_{21}$ are massless, $\mu(N_{12})=\mu(N_{21})=0$. The mion "masses" (5.7),(5.8) are not the pole masses but simply the low energy values of mass terms in their propagators, the only pole masses are given in (5.2).\\

5.3 \,\,\,   $\rm\bf br2$ vacua, $\bd/N_F\ll 1$\\

Instead of (5.2), the pole mass of mions is parametrically smaller now, see (1.17),(4.4),
\bq
\mu^{\rm pole}(M)=\frac{Z_q^{\, 2}\la^2}{z_M(\la,\mu^{\rm pole}(M))}\sim Z_q^{\, 2}\la\Bigl (\frac{\la}{\mph}\Bigr )^{\frac{N_F}{3(2N_c-N_F)}},\quad \frac{\mu^{\rm pole}(M)}{\mu^{\rm pole}_2(\Phi)}
\sim Z_q^{\, 2}\ll 1\,,
\eq
while instead of (5.3) we have now, see (4.5),
\bq
\mu_{q,2}=\frac{\langle M_2\rangle}{Z_q\la}\sim\frac{m_Q\mph}{Z_q\la},\quad {\tilde\mu}^{\rm pole}_{q,2}\sim\frac{\la}
{Z_q}\Bigl (\frac{m_Q\mph}{\la^2}\Bigr )^{N_F/3\nd}\gg\mu^{\rm pole}_{q,1}\,,
\eq
\bq
{\ov\mu}_{\rm gl,1}\sim \la\Bigl (\frac{\langle N_1\rangle}{\la^2}\Bigr )^{N_F/3N_c}\sim Z_q^{1/2}\la\Bigl(\frac{m_Q}{\la}
\Bigr )^{N_F/3N_c}\gg{\ov\mu}_{\rm gl,2}\,,\quad \frac{{\ov\mu}_{\rm gl,1}}{m^{\rm pole}_{Q,1}}\sim Z_q^{1/2}\ll 1\,,
\eq
\bq
\quad\frac{{\ov\mu}_{\rm gl,1}}{{\tilde\mu}^{\rm pole}_{q,2}}\sim Z_q^{3/2}\Bigl(\frac{\mo}{\mph}\Bigr )^{N_F/3\nd}\gg 1\,,\quad \la\ll {\mph}\ll Z_q^{\,3/2}\mo\,,\quad Z_q\sim\exp\{-\frac{\nd}{7\bd}\}\ll 1\,.
\eq

Hence, at the condition (5.13), the largest mass is ${\ov\mu}_{\rm gl,1}$ and the overall phase is also $Higgs_1-HQ_2$. But now, at $\bd/\nd\ll 1$, it looks unnatural to require $\bd^{\, \prime}=(\bd-2n_1)>0$. Therefore, with $n_1/\nd=O(1)$, the lower energy theory at $\mu<{\ov\mu}_{\rm gl,1}$ has $\bd^{\, \prime}<0$ and is in the logarithmic IR free regime in the range of scales $\qtp<\mu<{\ov\mu}_{\rm gl,1}$. Then instead of (5.5) (ignoring all logarithmic renormalization factors),
\bq
\lym^{(\rm br2)}\ll\qtp\sim \frac{\langle M_2\rangle_{\rm br2}}{\la}\frac{1}{z_q(\la,\muo)}\sim\frac{\mph}{Z_q^
{3/2}\mo}\,{\ov\mu}_{\rm gl,1}\ll {\ov\mu}_{\rm gl,1}\,.
\eq

The Lagrangian of mions and nions has now the form (5.6) with a replacement\\ $z^{\,\prime}_M (\muo,\qtp)\sim 1$ and so $\mu(M_{22})\sim\mu(M_{11})\sim\mu(M_{12})\sim\mu(M_{21})$ now, see (5.7),(5.8),(5.13),
\bq
\mu(M_{ij})\sim\frac{Z_q^2\la^2}{z_M(\la,\muo)\mph}\sim Z_q^{3/2}\Bigl (\frac{\mo}{\mph}\Bigr )\muo\gg\muo\,,
\eq
while, instead of (5.9), the mass of nions looks now as
\bq
\mu^{\rm pole}(N_{11})\sim \frac{\mph\langle N_1\rangle_{\rm br2}}{z_q(\la,\muo)\la^2}\sim Z_q^{1/2}\Bigl (\frac{\mph}{\mo}\Bigr )\,\muo\,.
\eq

On the whole for the mass spectra in this case. a) The heaviest are $N_F^2$ mions with the pole masses (5.10) (the "masses" (5.15) are not the pole masses but simply the low energy values of mass terms in the mion propagators). \, b)
The next are the masses (5.12) of $n_1(2\nd-n_1)$ higgsed gluons and their superpartners. \, c) There is a large number
of flavored hadrons, mesons and baryons, made of non-relativistic and weakly confined (the string tension being $\sqrt{\sigma}\sim\lym^{(\rm br2)}\ll\qtp$\,) quarks $\odq_2, \dq_2$ with unhiggsed colors. The mass spectrum of low-lying flavored mesons is Coulomb-like with parametrically small mass differences, $\Delta\mu_H/\mu_H=O(\bd^{\,2}/N^2_F)\ll 1$.\, d) A large number of gluonia made of $SU(\nd-n_1)$ gluons with the mass scale $\sim\lym^{(\rm br2)}$.\, e) $n_1^2$ nions $N_{11}$ with the masses (5.16).\, f) The hybrid nions $N_{12}, N_{21}$ are massless.\\

\hspace*{1cm}{\bf 5.4 \,\,\,    Special} vacua,\,\, $n_1=\nd,\,\, n_2=N_c$\\

Nothing specific for the dynamical scenario $\# 1$ with the diquark condensate (DC) was used in a description of these vacua in section 9.3 of \cite{ch1}. Hence, all remains the same in the dynamical scenario $\# 2$ here.\\

\section{Direct theory. Broken flavor symmetry.\\{\hspace*{1.2cm}} The region $\mathbf{\mo\ll\mph\ll\la^2/m_Q}$ }

6.1 \,\,\,   $\rm\bf br1$ vacua\\

The values of quark condensates are here
\bq
\langle\qqo\rangle_{\rm br1}\simeq\frac{N_c}{N_c-n_1}\, m_Q\mph\,,\quad \langle\qqt\rangle_{\rm br1}\sim \la^2\Bigl (\frac{\la}{\mph}\Bigr )^{\frac{n_1}{N_c-n_1}}\Bigl (\frac{m_Q}{\la}\Bigr )^{\frac{n_2-N_c}{N_c-n_1}}\,,
\eq
\bq
\frac{\langle\qqt\rangle_{\rm br1}}{\langle\qqo\rangle_{\rm br1}}\sim\Bigl (\frac{\mo}{\mph}\Bigr )^{\frac{N_c}{N_c-n_1}}\ll 1\,.\nonumber
\eq
From these, the values of some potentially relevant masses look as
\bq
\mgo^2\sim \Bigl (a_*\sim 1\Bigr )z_Q(\la,\mgo)\langle\qqo\rangle_{\rm br1} \,,\quad z_Q(\la,\mgo)\sim
\Bigl (\frac{\mgo}{\la}\Bigr )^{\bo/N_F}\,,  \nonumber
\eq
\bq
\mgo\sim\la \Bigl (\frac{m_Q\mph}{\la^2}\Bigr )^{N_F/3\nd}\gg\mgt\,,
\eq
\bq
\langle m^{\rm tot}_{Q,2}\rangle=\frac{\langle\qqo\rangle_{\rm br1}}{\mph}\sim m_Q\,,\quad  {\tilde m}^{\rm pole}_{Q,2}=\frac{\langle m^{\rm tot}_{Q,2}\rangle_{\rm br1}}{z_Q(\la,{\tilde m}^{\rm pole}_{Q,2})}\,,\nonumber
\eq
\bq
{\tilde m}^{\rm pole}_{Q,2}\sim\la\Bigl (\frac{m_Q}{\la}\Bigr )^{N_F/3N_c}\gg m^{\rm pole}_{Q,1},\quad\,\,\, \frac{{\tilde m}^{\rm pole}_{Q,2}}{\mgo}\sim\Bigl (\frac{\mo}{\mph}\Bigr )^{N_F/3\nd}\ll 1.
\eq
Hence, the largest mass is $\mgo$ due to higgsing of ${\ov Q}_1, Q_1$ quarks and the overall phase is $Higgs_1-HQ_2$.

The lower energy theory at $\mu<\mgo$ has $N_c^\prime=N_c-n_1$ colors and $n_2\geq N_f/2$ flavors. At $2n_1<\bo$ it
remains in the conformal window with ${\rm b}_o^\prime>0$, while at $2n_1>\bo,\,\,{\rm b}_o^\prime<0$ it enters the logarithmic IR free perturbative regime.

We start with $\bo^\prime>0$. Then the value of the pole mass of quarks $\oqt,\,\sqt$ with unhiggsed colors looks as
\bq
m^{\rm pole}_{\sqt}=\frac{\langle m^{\rm tot}_{Q,2}\rangle_{\rm br1}}{z_Q(\la,\mgo)z_Q^{\,\prime}(\mgo,
m^{\rm pole}_{\sqt})}\,,\quad z_Q^{\,\prime}(\mgo,m^{\rm pole}_{\sqt})\sim\Bigl (\frac{m^{\rm pole}_{\sqt}}{\mgo}\Bigr )^{{\rm b}_o^\prime/n_2}\,,\nonumber
\eq
\bq
m^{\rm pole}_{\sqt}\sim\la\Bigl (\frac{\la}{\mph}\Bigr )^{\frac{n_1}{3(N_c-n_1)}}\Bigl (\frac{m_Q}{\la}\Bigr )^{\frac{n_2-n_1}{3(N_c-n_1)}}\sim\lym^{(\rm br1)}\,.
\eq
It is technically convenient to retain all fion fields $\Phi$ although, in essence, they are too heavy and dynamically irrelevant at $\mph\gg\mo$. After integrating out all heavy higgsed gluons and quarks ${\ov Q}_1, Q_1$, we write the Lagrangian at $\mu=\mu_{\rm gl,1}$  in the form
\bq
K=\Bigl [\,z_{\Phi}(\la,\mgo){\rm Tr}(\Phi^\dagger\Phi)+z_Q(\la,\mu^2_{\rm gl,1})\Bigl (K_{{\sq}_2}+K_{\Pi}\Bigr )
\,\Bigr ],\quad z_{\Phi}(\la,\mgo)=1/z^2_Q(\la,\mgo)\,,\nonumber
\eq
\bq
K_{{\sq}_2}={\rm Tr}\Bigl ({\sq}^{\dagger}_2 {\sq}_2 +({\sq}_2\ra
{\oq}_2 )\Bigr )\,, \quad K_{\Pi}= 2{\rm Tr}\sqrt{\Pi^{\dagger}_{11}\Pi_{11}}+K_{\rm hybr},
\eq
\bq
K_{\rm hybr}={\rm Tr}\Biggl (\Pi^{\dagger}_{12}\frac{1}{\sqrt{\Pi_{11}\Pi^{\dagger}_{11}}}\Pi_{12}+
\Pi_{21}\frac{1}{\sqrt{\Pi^{\dagger}_{11}\Pi_{11}}}\Pi^\dagger_{21}\Biggr ),\nonumber
\eq
\bq
W=\Bigl [-\frac{2\pi}{\alpha(\mu_{\rm gl,1})}{\textsf S}\Bigr ]+\frac{\mph}{2}\Biggl [{\rm Tr}\, (\Phi^2) -\frac{1}{\nd}\Bigl ({\rm Tr}\,\Phi\Bigr)^2\Biggr ]+{\rm Tr}\Bigl ({\oq_2}m^{\rm tot}_{{\sq}_2}{\sq}_2\Bigr )+W_{\Pi},\nonumber
\eq
\bq
W_{\Pi}= {\rm Tr}\Bigl (m_Q\Pi_{11}+m^{\rm tot}_{{\sq}_2}\,\Pi_{21}\frac{1}{\Pi_{11}}\Pi_{12}\Bigr )-
{\rm Tr}\Bigl (\Phi_{11}\Pi_{11}+\Phi_{12}\Pi_{21}+\Phi_{21}\Pi_{12} \Bigr ),
\quad m^{\rm tot}_{{\sq}_2}=(m_Q-\Phi_{22}).\nonumber
\eq
In (6.5): $\oqt,\, \sqt$ and $\textsf V$ are the active ${\ov Q}_2, Q_2$ guarks and gluons with unhiggsed colors ($\textsf S$ is their field strength squared), $\Pi_{12}, \Pi_{21}$ are the hybrid pions (in essence, these are the quarks ${\ov Q}_2, Q_2$ with higgsed colors), $z_Q(\la,\mu^2_{\rm gl,1})$ is the corresponding perturbative renormalization factor of massless quarks, see (6.2), while $z_{\Phi}(\la,\mgo)$ is that of fions. Evolving now down in the scale and integrating out at $\mu<\lym^{(\rm br1)}$ quarks  $\oq_2,\, \sq_2$ as heavy ones and unhiggsed gluons, we obtain the Lagrangian of pions and fions
\bq
K=\Bigl [z_{\Phi}(\la,\mgo){\rm Tr}\Bigl(\Phi^\dagger_{11}\Phi_{11}+\Phi^\dagger_{12}\Phi_{12}+\Phi^\dagger_{21}
\Phi_{21}+z^{\,\prime}_{\Phi}(\mgo,m^{\rm pole}_{\sqt})\Phi^\dagger_{22}\Phi_{22}\Bigr )+z_Q(\la,\mu^2_{\rm gl,1})K_{\Pi}\Bigr ],\,\nonumber
\eq
\bq
W=(N_c-n_1)S+W_{\Phi}+W_{\Pi}\,,\quad S=\Biggl [\frac{\la^{\bo}\det m^{\rm tot}_{{\sq}_2}}{\det \Pi_{11}}\Biggr ]^{\frac{1}{N_c-n_1}}\,,
\eq
\bq
W_{\Phi}=\frac{\mph}{2}\Biggl [{\rm Tr} (\Phi^2) -\frac{1}{\nd}\Bigl ({\rm Tr}\,\Phi\Bigr)^2\Biggr ],\quad
z^{\,\prime}_{\Phi}(\mgo,m^{\rm pole}_{\sqt})\sim\Bigl (\frac{\mgo}{m^{\rm pole}_{\sqt}}\Bigr )^{2{\rm b}_o^\prime/n_2}\,.
\nonumber
\eq
We obtain from (6.6) that all fions are heavy with the "masses"
\bq
\mu(\Phi_{11})\sim\mu(\Phi_{12})\sim\mu(\Phi_{21})\sim\frac{\mph}{z_{\Phi}(\la,\mgo)}\sim\Bigl (\frac{\mph}{\mo}\Bigr )^{N_c/\nd}\mgo\gg\mgo\,,
\eq
\bq
\mu(\Phi_{22})\sim\frac{\mph}{z_{\Phi}(\la,\mgo)z^{\,\prime}_{\Phi}(\mgo,m^{\rm pole}_{\sqt})}\sim
\Bigl (\frac{\mph}{\mo}\Bigr )^{\frac{N_c}{N_c-n_1}}\, m^{\rm pole}_{\sqt}\gg m^{\rm pole}_{\sqt}\,.
\eq
These are not the pole masses but simply the low energy values of mass terms in their propagators. All fions are dynamically irrelevant at all scales $\mu<\la$. The mixings of $\Phi_{12}\leftrightarrow\Pi_{12},\, \Phi_{21}
\leftrightarrow\Pi_{21}$ and $\Phi_{11}\leftrightarrow\Pi_{11}$ are parametrically small and are neglected. We
obtain then for the masses of pions $\Pi_{11}$
\bq
\mu(\Pi_{11})\sim\Bigl (\frac{\mo}{\mph}\Bigr )^{\frac{N_c(\bo-2n_1)}{3\nd(N_c-n_1)}}\,\lym^{(\rm br1)}\sim\Bigl (\frac{\mo}{\mph}\Bigr )^{\frac{N_c(\bo-2n_1)}{3\nd(N_c-n_1)}}\, m^{\rm pole}_{\sqt}\ll m^{\rm pole}_{\sqt}\,,
\eq
and, finally, the hybrids $\Pi_{12}, \Pi_{21}$ are massless, $\mu(\Pi_{12})=\mu(\Pi_{21})=0$.

At $2n_1>\bo$ the RG evolution at $m^{\rm pole}_{\sqt}<\mu<\mgo$ is only slow logarithmic (and is neglected). We replace then $z^{\,\prime}_Q(\mgo,m^{\rm pole}_{\sqt})\sim 1$ in (6.4) and $z^{\,\prime}_{\Phi}(\mgo,m^{\rm pole}_{\sqt})\sim 1$ in (6.8) and obtain
\bq
\mu(\Phi_{22})\sim\mu(\Phi_{11})\sim\Bigl (\frac{\mph}{\mo}\Bigr )^{N_c/\nd}\mgo\gg\mgo\,,
\eq
\bq
\mu(\Pi_{11})\sim m^{\rm pole}_{\sqt}\sim\frac{m_Q}{z_Q(\la,\mu^2_{\rm gl,1})}\sim\la\Bigl (\frac{\la}{\mph}\Bigr )^{\bo/3\nd}\Bigl (\frac{m_Q}{\la}\Bigr )^{2\,\bd/3\nd}\sim\Bigl (\frac{\mo}{\mph}\Bigr )^{N_c/\nd}\mgo\ll\mgo.\nonumber
\eq
\bq
\frac{\lym^{(\rm br1)}}{m^{\rm pole}_{\sqt}}\sim\Bigl (\frac{\mo}{\mph}\Bigr )^{\Delta}\ll 1,\quad
\Delta=\frac{N_c(2n_1-\bo)}{3\nd(N_c-n_1)}>0\,.
\eq
\newpage

6.2 \,\,\,   $\rm\bf br2$ and $\rm \bf special$ vacua\\

At $n_2<N_c$ there are also $\rm br2$ - vacua. All their properties can be obtained by a replacement $n_1\leftrightarrow n_2$ in formulas of the preceding section 6.1. The only difference is that, because $n_2\geq N_F/2$ and so $2n_2>\bo$, there is no analog of the conformal regime at $\mu<\mu_{\rm gl,1}$ with $2n_1<\bo$. I.e. at $\mu<\mu_{\rm gl,2}$ the lower energy theory will be always in the perturbative IR free logarithmic regime and the overall phase will be $Higgs_2-HQ_1$.

As for the special vacua, all their properties can also be obtained with $n_1=\nd,\, n_2=N_c$
in formulas of the preceding section 6.1.\\

\section{Dual theory. Broken flavor symmetry.\\{\hspace*{1.2cm}} The region $\mathbf{\mo\ll\mph\ll\la^2/m_Q}$ }

\hspace*{1cm}{\bf 7.1 \,\,\,   $\rm\bf br1$} vacua, $\,\,\bd/N_F\ll 1$\\

We recall, see (1.12), that condensates of mions and dual quarks in these vacua are
\bq
\langle M_1\rangle_{\rm br1}\simeq\frac{N_c}{N_c-n_1}\, m_Q\mph\,,\quad \langle M_2\rangle_{\rm br1}\sim \la^2\Bigl (\frac{\la}{\mph}\Bigr )^{\frac{n_1}{N_c-n_1}}\Bigl (\frac{m_Q}{\la}\Bigr )^{\frac{n_2-N_c}{N_c-n_1}}\,,
\eq
\bq
\frac{\langle M_2\rangle_{\rm br1}}{\langle M_1\rangle_{\rm br1}}\sim\Bigl (\frac{\mo}{\mph}\Bigr )^{\frac{N_c}{N_c-n_1}}\ll 1\,,\nonumber
\eq
\bq
\langle N_2\rangle_{\rm br1}\equiv\langle {\ov q}_2 q_2(\mu=\la)\rangle_{\rm br1}=Z_q\frac{\langle M_1\rangle_{\rm br1}\la}{\mph}\sim Z_q m_Q\la\gg\langle N_1\rangle_{\rm br1}\,,\nonumber
\eq
and some potentially relevant masses look here as
\bq
\qo=\qo(\mu=\la)=\frac{\langle M_1\rangle_{\rm br1}}{Z_q\la}\sim\frac{m_Q\mph}{Z_q\la}\,,\quad
\frac{\qt}{\qo}= \frac{\langle M_2\rangle_{\rm br1}}{\langle M_1\rangle_{\rm br1}}\ll 1\,,
\eq
\bq
Z_q\sim\exp \Bigl\{-\frac{1}{3{\ov a}_{*}}\Bigr\}\sim \exp \Bigl\{-\frac{\nd}{7\bd}\Bigr\}\ll 1\,,\nonumber
\eq
\bq
\qop\sim \frac{\la}{Z_q}\Bigl (\frac{m_Q\mph}{\la^2}\Bigr )^{N_F/3\nd}\gg\qtp\,,\quad
\frac{\lym^{(\rm br1)}}{\qop}\sim Z_q\Bigl (\frac{\mo}{\mph}\Bigr )^{\frac{n_2 N_c}{3\nd(N_c-n_1)}}\ll 1\,,
\eq
\bq
\mut\sim\la\Bigl (\frac{\langle N_2\rangle}{\la^2}\Bigr )^{N_F/3N_c}\sim Z_q^{1/2}\la
\Bigl (\frac{m_Q}{\la}\Bigr )^{N_F/3N_c}\gg\muo\,,\nonumber
\eq
\bq
\frac{\mut}{\qop}\sim Z_q^{3/2}\Bigl (\frac{\mo}{\mph}\Bigr )^{N_F/3\nd}\ll 1\,.
\eq
Hence, the largest mass is $\qop$ while the overall phase is $HQ_1-HQ_2$. We consider below only the case $n_1<\bo/2$, so that the lower energy theory with $\nd$ colors and $N^\prime_F=n_2$ flavors at $\mu<\qop$ remains in the conformal
window.

After integrating out the heaviest quarks ${\ov q}_1, q_1$ at $\mu<\qop$ and ${\ov q}_2, q_2$ quarks at $\mu<\qtp$ and, finally, all $SU(\nd)$ dual gluons at $\mu<\lym^{(\rm br1)}$, the Lagrangian of mions looks as
\bq
K=\frac{z_M(\la,\qop)}{Z^2_q\la^2}\,{\rm Tr}\Bigl [\, M_{11}^\dagger M_{11}+M_{12}^\dagger M_{12}+M_{21}^\dagger M_{21}+z^{\,\prime}_{M}(\qop,\qtp) M_{22}^\dagger M_{22} \Bigr ]\,,
\eq
\bq
W=-\nd S+W_M\,,\quad\quad S=\Bigl (\frac{\det M}{\la^{\bo}}\Bigr )^{1/\nd}\,,
\quad \lym^{(\rm br1)}=\langle S\rangle^{1/3}\sim\Bigl (m_Q\langle M_2\rangle\Bigr )^{1/3}\,.\nonumber
\eq
\bq
W_M=m_Q{\rm Tr} M-\frac{1}{2\mph}\Bigl [\,{\rm Tr}(M^2)-\frac{1}{N_c}({\rm Tr} M)^2 \Bigr ]\,,
\quad z_M(\la,\qop)\sim \Bigl (\frac{\la}{\qop}\Bigr )^{2\,\bd/N_F}\gg 1\,.\nonumber
\eq

From (7.5)\,: the hybrids $M_{12}$ and $M_{21}$ are massless, $\mu(M_{12})=\mu(M_{21})=0$, while the pole mass of $M_{11}$ is (compare with (6.9)\,)
\bq
\mu^{\rm pole}(M_{11})\sim\frac{Z^2_q\la^2}{z_M(\la,\qop)\mph}\,,\quad \frac{\mu^{\rm pole}(M_{11})}{\lym^{(\rm br1)}}
\sim Z^2_q\Bigl (\frac{\mo}{\mph}\Bigr )^{\frac{N_c(\bo-2n_1)}{3\nd(N_c-n_1)}}\ll 1\,.
\eq

The parametric behavior of $\qtp$ and $z^{\,\prime}_{M}(\qop,\qtp)$ depends on the value $\mph\lessgtr{\tilde\mu}_{\Phi,1}$ (see below). We consider first the case $\mph\gg{\tilde\mu}_{\Phi,1}$ so that, by definition, the lower energy theory with $\nd$ colors and $n_2$ flavors had enough "time" to evolve and entered already the new conformal regime at $\qtp<\mu\ll\qop$, with ${\rm\ov b\,}^\prime_o/\nd=(3\nd-n_2)/\nd=O(1)$ and ${\ov a\,}^\prime_*=O(1)$. Hence, when the quarks ${\ov q}_2, q_2$ decouple as heavy ones at $\mu<\qtp$, the coupling ${\ov a}_{YM}$ of the remained $SU(\nd)$ Yang-Mills theory is ${\ov a}_{YM}\sim {\ov a\,}^\prime_*=O(1)$ and this means that $\qtp\sim\lym^{(\rm br1)}$. This can be obtained also in a direct way. The running mass of quarks ${\ov q}_2, q_2$ at $\mu=\qop$ is, see (7.1)-(7.3),
\bq
\mu_{q,2}(\mu=\qop)=\frac{\langle M_2\rangle_{\rm br1}}{\langle M_1\rangle_{\rm br1}}\,\qop\,,\quad \qtp=\frac{\mu_{q,2}
(\mu=\qop)}{z^{\,\prime}_q(\qop,\qtp)}\sim \lym^{(\rm br1)}\sim\Bigl (m_Q\langle M_2\rangle\Bigr )^{1/3}\,,
\eq
\bq
z^{\,\prime}_q(\qop,\qtp)=\Bigl (\frac{\qtp}{\qop}\Bigr )^{\frac{{\rm\ov b\,}^\prime_o}{n_2}}\rho\,,\quad \rho=\Bigl (\frac{{\ov a\,}_*}{{\ov a\,}^\prime_*}\Bigr )^{\frac{\nd}{n_2}}\exp\Bigl\{\frac{\nd}{n_2}\Bigl (\frac{1}{{\ov a\,}_*}-\frac{1}{{\ov a\,}^\prime_*}\Bigr )\Bigr \}
\sim \exp\Bigl\{\frac{\nd}{n_2}\frac{1}{{\ov a\,}_*}\Bigr\}\gg 1\,.\nonumber
\eq
We obtain from (7.5) that the main contribution to the mass of mions $M_{22}$ originates from the non-perturbative term
$\sim S$ in the superpotential and, using (7.5),(7.7),
\bq
z^{\,\prime}_{M}(\qop,\qtp)=\frac{{\ov a}_f(\mu=\qop)}{{\ov a}_f(\mu=\qtp)}\Bigl (\frac{1}{z^{\,\prime}_q(\qop,\qtp)}
\Bigr )^2\sim\Bigl (\frac{1}{z^{\,\prime}_q(\qop,\qtp)}\Bigr )^2\,,
\eq
\bq
\mu(M_{22})\sim\frac{Z_q^2\la^2}{z_M(\la,\qop) z^{\,\prime}_{M}(\qop,\qtp)}\Biggl (\frac{\langle S\rangle}{\langle M_2\rangle^2}=\frac{\langle M_1\rangle}{\langle M_2\rangle}\frac{1}{\mph}\Biggr )_{\rm br1}\,\,\sim \lym^{(\rm br1)}\sim\qtp\,.
\eq

We consider now the region $\mo\ll\mph\ll{\tilde\mu}_{\Phi,1},\, 2n_1\lessgtr\bo$ where, by definition, $\qtp$ is too close to $\qop$, so that the range of scales $\qtp<\mu<\qop$ is too small and the lower energy theory at $\mu<\qop$ has no enough "time" to enter a new regime (conformal at $2n_1<\bo$ or strong coupling one at $2n_1>\bo$) and remains in the weak coupling logarithmic regime. Then, ignoring logarithmic effects in renormalization factors, $z^{\,\prime}_q(\qop,\qtp)\sim z^{\,\prime}_{M}(\qop,\qtp)\sim 1$, and keeping as always only the exponential dependence on $\nd/\bd$\,:
\bq
\qtp\sim\frac{\langle M_2\rangle_{\rm br1}}{\langle M_1\rangle_{\rm br1}}\,\qop\,,\quad\quad \frac{\lym^{(\rm br1)}}
{\qtp}\ll 1\quad\ra\quad \mo\ll\mph\ll{\tilde\mu}_{\Phi,1}\,,\nonumber
\eq
\bq
{\tilde\mu}_{\Phi,1}\sim\exp\Bigl\{\frac{(N_c-n_1)}{2n_1}\frac{1}{{\ov a\,}_*}\Bigr\}\mo \gg \mo\,.
\eq

The pole mass of mions $M_{22}$ looks in this case as
\bq
\frac{\mu^{\rm pole}(M_{22})}{\mu^{\rm pole}(M_{11})}\sim\frac{\langle M_1\rangle_{\rm br1}}{\langle M_2\rangle_{\rm br1}}\gg 1,\quad\frac{\mu^{\rm pole}(M_{22})}{\lym^{(\rm br1)}}\sim Z^2_q\Bigl (\frac{\mph}{\mo}\Bigr )^{\frac{2n_1 N_c}{3\nd(N_c-n_1)}}\ll 1\,.
\eq

On the whole, see (7.10), the mass spectrum at $\mo\ll\mph\ll{\tilde\mu}_{\Phi,1}$ and $2n_1\lessgtr\bo$ looks as follows. a) There is a large number of heaviest hadrons made of weakly coupled (and weakly confined, the string tension being $\sqrt{\sigma}\sim\lym^{(\rm br1)}\ll\qop)$ nonrelativistic quarks ${\ov q}_1, q_1$, the scale of their masses is $\qop$, see (7.3).\, b) The next physical mass scale is due to $\qtp,\,\, \lym^{(\rm br1)}\ll\qtp\ll\qop$. Hence, there is also a large number of hadrons made of weakly coupled and weakly confined nonrelativistic quarks ${\ov q}_2, q_2$, the scale of their masses is $\qtp$, see (7.10), and a large number of heavy hybrid hadrons with the masses $\sim (\qop+\qtp)$.
Because all quarks are weakly coupled and non-relativistic in all three flavor sectors, $"11",\, "22"$ and $"12+21"$, the mass spectrum of low-lying flavored mesons is Coulomb-like with parametrically small mass differences $\Delta\mu_H/\mu_H=O(\bd^2/\nd^2)\ll 1$.\, c) A large number of gluonia made of $SU(\nd)$ gluons, with the mass scale $\sim\lym^{(\rm br1)}\sim\Bigl (m_Q\langle M_2\rangle\Bigr )^{1/3}$, see (7.5),(7.1).\,    d) $n^2_2$ mions $M_{22}$ with the pole masses $\mu^{\rm pole}(M_{22})\ll\lym^{(\rm br1)}$, see (7.11).\, e) $n^2_1$ mions $M_{11}$ with the pole masses $\mu^{\rm pole}(M_{11})\ll\mu^{\rm pole}(M_{22})$, see (7.6),(7.11).\, f) $2n_1n_2$ hybrids $M_{12}, M_{21}$ are massless, $\mu(M_{12})=\mu(M_{21})=0$.

The pole mass of quarks ${\ov q}_2, q_2$ is smaller at ${\tilde\mu}_{\Phi,1}\ll\mph\ll \la^2/m_Q$ and $2n_1<\bo$, and stays at $\qtp\sim\lym^{(\rm br1)}$, while the mass of mions $M_{22}$ is larger and also stays at $\mu(M_{22})\sim\lym^{(\rm br1)}$.\\

7.2 \,\,\,   $\rm\bf br2$ and $\rm \bf special$ vacua, $\bd/N_F\ll 1$\\

The condensates of mions look in these br2 - vacua as in (7.1) with the exchange $1\leftrightarrow 2$. The largest mass  is $\qtp$,
\bq
\qtp\sim \frac{\la}{Z_q}\Bigl (\frac{m_Q\mph}{\la^2}\Bigr )^{N_F/3\nd}\gg\qop\,,\quad
\frac{\lym^{(\rm br2)}}{\qtp}\sim Z_q\Bigl (\frac{\mo}{\mph}\Bigr )^{\frac{n_1 N_c}{3\nd(N_c-n_2)}}\ll 1\,,
\eq
and the overall phase is $HQ_1-HQ_2$. After decoupling the heaviest quarks ${\ov q}_2, q_2$ at $\mu<\qtp$ the lower energy theory remains in the weak coupling logarithmic regime at, see (7.10),
\bq
\frac{\lym^{(\rm br2)}}{\qop}\ll 1 \quad\ra\quad \mo\ll\mph\ll {\tilde\mu}_{\Phi,2}\,,\quad
\frac{{\tilde\mu}_{\Phi,2}}{\mo}\sim\exp\Bigl\{\frac{(N_c-n_2)}{2n_2}\frac{1}{{\ov a\,}_*}\Bigr\}\gg 1\,.
\eq
Hence, the  mass spectra in this range $\mo<\mph\ll {\tilde\mu}_{\Phi,2}$ can be obtained from corresponding formulas in section 7.1 by the replacements $n_1\leftrightarrow n_2$.

But because $n_2\geq N_F/2$, the lower energy theory with $1<n_1/\nd<3/2$ is in the strong coupling regime at $\mph\gg{\tilde\mu}_{\Phi,2}$, with ${\ov a}(\mu)\gg 1$ at $\lym^{(\rm br2)}\ll\mu\ll\qtp$. We do not consider the strong coupling regime in this paper. \\

As for the special vacua, the overall phase is also $HQ_1-HQ_2$ therein.  The mass spectra are obtained by substituting $n_1=\nd$ into the formulas of section 7.1. At $5/3<N_F/N_c<2$ and $\mph\gg{\tilde\mu}_{\Phi,1}$ the lower energy theory in these special vacua enters the strong coupling regime at $\lym^{(\rm spec)}\ll\mu\ll\qop$.

\section{Conclusions}

\hspace*{0.5cm}The mass spectra of the direct $\bf \Phi$ -theory and its dual variant, the $\bf d\Phi$ -theory, were calculated in \cite{ch1} within the dynamical scenario $\#1$ which implies a possibility of (quasi)spontaneous breaking of chiral flavor symmetry.  I.e., the phase is formed with a large coherent diquark condensate (DC), $\langle{\ov Q} Q\rangle$, and without higgsing of quarks, $\langle{\ov Q}\rangle=\langle Q\rangle=0$. As a result, the light quarks acquire the  large dynamical constituent mass $\mu_C\sim\langle{\ov Q} Q\rangle^{1/2}$ and much lighter (pseudo) Nambu-Goldstone pions appear. This paper continues \cite{ch1}. The mass spectra of the $\bf \Phi$ and $\bf d\Phi$ theories are calculated here within the dynamical scenario $\#2$ which implies that the DC phase is not formed and the quarks may be in the two different phases only\,: either they are in the HQ (heavy quark) phase where they are confined, or they are higgsed at appropriate values of the lagrangian parameters.

As was shown above in the main text, the use of the additional small parameter $\bd/N_F\ll 1$ allows to trace explicitly
{\it parametrical} differences in the mass spectra of the direct $\bf \Phi$ and dual $\bf d\Phi$ theories. This shows that these two theories are not equivalent. A similar situation takes place when comparing the mass spectra of the ordinary ${\cal N}=1$ SQCD and its dual variant \cite{ch2}.

At present, unfortunately, to calculate the mass spectra of ${\cal N}=1$ SQCD-like theories one is forced to assume a definite dynamical scenario. But nevertheless, from our standpoint, a most important thing at present is a very ability to calculate the mass spectra of various ${\cal N}=1$ SQCD-like theories, even within a given dynamical scenario. It seems clear that further developments will allow to find a unique right scenario in each such theory.\\

This work is supported in part by  Ministry of Education and Science of the Russian Federation and RFBR grant 12-02-00106-a.


\vspace*{2mm}

\begin{thebibliography}{99}
\bibitem{ch1}
V.L. Chernyak, Mass spectrum in SQCD with additional fields. I, arXiv:\,1205.0410[hep-th]
\bibitem{ch2}
V.L. Chernyak, On mass spectrum in SQCD and problems with the Seiberg duality.\\
Another scenario, JETP {\bf 114} (2012) 61,  arXiv:\, 0811.4283 [hep-th]
\bibitem{NSVZ1}
V. Novikov, M. Shifman, A. Vainshtein, V. Zakharov, Exact Gell-Mann-Low function of \\
supersymmetric Yang-Mills theories from instanton calculus, Nucl. Phys. {\bf B 229} (1983) 381
\bibitem{NSVZ2}
M. Shifman, A. Vainshtein, Solution of the anomaly puzzle in SUSY gauge theories and the Wilson
operator expansion, Nucl. Phys. {\bf B 277} (1986) 456
\bibitem{Konishi}
K. Konishi, Anomalous supersymmetry transformation of some composite operators in SQCD,
Phys. Lett. {\bf B 135} (1984) 439
\bibitem{S1}
N. Seiberg, Exact results on the space of vacua of four-dimensional SUSY gauge theories,
Phys. Rev. {\bf D 49} (1994) 6857,\,  hep-th/9402044
\bibitem{S2}
N. Seiberg, Electric - magnetic duality in supersymmetric nonabelian gauge theories,\\
Nucl. Phys. {\bf B 435} (1995) 129,\,  hep-th/9411149
\bibitem{ch3}
V.L. Chernyak, On mass spectrum in SQCD and problems with the Seiberg duality.\\
Equal quark masses, JETP {\bf 110} (2010) 383,  arXiv\,:\,0712.3167\, [hep-th]
\bibitem{ch4}
V.L. Chernyak, On mass spectrum in SQCD. Unequal quark masses, JETP {\bf 111} (2010) 949,
arXiv\,:\,0805.2299\, [hep-th]
\bibitem{VY}
G. Veneziano, S. Yankielowicz, An effective Lagrangian for the pure ${\cal N}=1$ supersymmetric Yang-Mills theory,
Phys. Lett. {\bf B 113} (1982) 231
\bibitem{KSV}
I. Kogan, M. Shifman, A. Vainshtein, Phys. Rev. {\bf D 53} (1996)
4526,\\  hep-th/9507170\,;\,\,\,  {\rm Err}: Phys. Rev. {\bf D 59} (1999) 109903
\end{thebibliography}
\end{document}